\documentclass[pre,showpacs,preprintnumbers,amsmath,amssymb,preprint]{revtex4}

\usepackage{graphicx}
\usepackage{dcolumn}
\usepackage{bm}

\usepackage[T1]{fontenc}
\usepackage[cp852]{inputenc}
\usepackage[a4paper]{geometry}
\usepackage{float}
\usepackage{units}
\usepackage{textcomp}
\usepackage{amsmath}
\usepackage{amssymb}
\usepackage{esint}
\usepackage{amsmath}    
\usepackage[all]{xy}
\usepackage{multirow, array}
 \usepackage[titletoc]{appendix}
\usepackage{rotating}

\DeclareGraphicsExtensions{.jpg}

\begin{document}

\title{Diffusion behavior in Nickel-Aluminium and Aluminium-Uranium diluted alloys}

 \author{Viviana P. Ramunni}
 \affiliation {CONICET - Avda. Rivadavia 1917, Cdad. de Buenos Aires, C.P. 1033, Argentina.}  
 \affiliation {Departamento de Materiales, CAC-CNEA, Avda. General Paz 1499, 1650 San Mart{\'i}n, Argentina.}
 \email{vpram@cnea.gov.ar}  
\thanks{This work was partially financed by CONICET - PIP 00965/2010.}
\date{\today}

\begin{abstract}

Impurity diffusion coefficients are entirely obtained from a low cost classical molecular statics technique (CMST). In particular, we show how the CMST is appropriate in order to describe the impurity diffusion behavior mediated by a vacancy mechanism. In the context of the five-frequency model, CMST allows to calculate all the microscopic parameters, namely: the free energy of vacancy formation, the vacancy-solute binding energy and the involved jump frequencies, from them, we obtain the macroscopic transport magnitudes such as: correlation factor, solvent-enhancement factor, Onsager and diffusion coefficients. Also, we report for the first time the behavior of diffusion coefficients for the solute-vacancy paired specie. We perform our calculations in diluted $NiAl$ and $AlU$ f.c.c. alloys. Our results are in perfect agreement with available experimental data for both systems and predict that for $NiAl$ the solute diffuses through a vacancy interchange mechanism, while for the $AlU$ system, a vacancy drag mechanism occurs \\ 

PACS number(s): Diffusion, Numerical Calculations, Vacancy mechanism, diluted Alloys, NiAl and AlU systems. 

\end{abstract}
\maketitle

\section{Introduction}

The low enrichment of $U-$Mo alloy dispersed in an $Al$ matrix is a prototype for new experimental nuclear fuels \cite{WEB}. When these metals are brought into contact, diffusion in the $Al/U-Mo$ interface gives rise to interaction phases. Also, when subjected to temperature and neutron radiation, phase transformation from $\gamma U$ to $\alpha U$ occurs and intermetallic phases develop in the U$-$Mo$/$Al interaction zone. Fission gas pores nucleate in these new phases during service producing swelling and deteriorating the alloy properties \cite{WEB,SAV05}. An important technological goal is to delay or directly avoid undesirable phase formation by inhibiting interdiffusion of $Al$ and $U$ components. Some of these compounds are believed to be responsible for degradation of properties \cite{MIR09}. On the other hand, there is an experimental work \cite{HOU71}, that argues that these undesirable phases have not influence on the mobility of $U$ in $Al$, based on the results of the effective diffusion coefficients calculated from the best fit of their permeation experimental curves. 

Another technique to study the diffusion of Uranium into Aluminum was based on the maximum rate of penetration of uranium into aluminum as function of the temperature \cite{BIE55}. From this perspective, the authors also report the activation energy values of Uranium mobility. In avoiding interdiffusion, Brossa et. al. \cite{BRO63} studied the efficient diffusion barriers that should have a good bonding effect and exhibit a good thermal conductivity at the same time. In this work, deposition methods have been developed and the diffusion behavior of the respective couples and triplets has been evaluated by metallographic, micro-hardness and electron microprobe analyses. The practical interest of a nickel barrier is shown by several publications concerning to the diffusion in the systems $AlNi$, $NiU$ and $AlNiU$. The knowledge of the binary system is the only satisfactory basis for the study of the ternary system, these binary systems are treated briefly before proceeding to the ternary. The study of the $NiAl$ binary system was, limited to solid samples of the sandwich-type, clamped together by a titanium screw and diffusion treatments have been carried out. Results from this work, have inspired as to also study the $NiAl$ together with the $AlU$ system. 

Therefore it is important to watch carefully and with special attention the initial microscopic processes that originate these intermetallic phases. In order to deal with this problem we started studying numerically the static and dynamic properties of vacancies and interstitials defects in the $Al$($U$) bulk and in the neighborhood of a $(111)Al/(001)\alpha U$ interface using molecular dynamics calculations \cite{PAS11,RAM10}. Here, we review our previous works \cite{PAS11,RAM10}, performing calculation of three diffusion coefficients, namely: the solvent self diffusion coefficient, the solute tracer diffusion coefficient and one more, never before calculated in the literature for this alloy, of the Uranium-vacancy paired specie. With this purpose we use analytical expressions of the diffusion parameters in terms of microscopical magnitudes. We have summarized the theoretical tools needed to express the diffusion coefficients in terms of microscopic magnitudes as, the jump frequencies, the free vacancy formation energy and the vacancy-solute binding energy. Then we starts with non-equilibrium thermodynamics in order to relate the diffusion coefficients with the phenomenological $L$-coefficients. The microscopic kinetic theory, allows us to write the Onsager coefficients in term of the jump frequency rates. At this point we follow the procedure of Okamura and Allnat \cite{OKA83}, and Allnatt and Lidiard \cite{ALL03}. 

The jump frequencies are identified by the model developed further by Le Claire in Ref. \cite{LEC78}, known as the five-frequency model for f.c.c lattices. The method includes the jump frequency associated with the migration of the host atom in the presence of an impurity at a first nearest neighbor position. All this concepts need to be put together in order to correctly describe the diffusion mechanism. Hence, in the context of the shell approximation, we follow the technique in Ref. \cite{ALL03} to obtain the corresponding transport coefficients which are related to the diffusion coefficients through the flux equations. A similar procedure for f.c.c. structures was performed by Mantina et al. \cite{MAN09} for $Mg$, $Si$ and $Cu$ diluted in $Al$ but using density functional theory (DFT). Also, for b.c.c. structures, Choudhury \textit{et al.} \cite{CHO11} have calculated the self-diffusion and solute diffusion coefficients in diluted $\alpha FeNi$ and $\alpha FeCr$ alloys including an extensive analysis of the phenomenological $L$-coefficients using DFT calculations. Also the authors discuss about the risk induced by radiation on based $FeNi$ and $FeCr$ alloys.  

In the present work, we do not employ DFT, instead we use a classical molecular statics technique, the Monomer method \cite{RAM06}. This much less computationally expensive method allows us to compute at low cost a bunch of jump frequencies from which we can perform averages in order to obtain more accurate effective frequencies. Also, for the first time in the literature, we have calculated the diffusion coefficient of the paired solute-vacancy specie by exploring all the possibilities of the solute mobility, either via direct exchange solute-vacancy mechanism or by a vacancy drag mechanism in which the solute-vacancy pair migrates as a complex defect. Although we use classical methods, we reproduce the migration barriers for $NiAl$ using the SIESTA code coupled to the Monomer method \cite{RAM09} using pipes of UNIX for the communications. 

We proceed as follows, first of all we validate the five-frequency model using the $NiAl$ system as a reference case for which there are a large amount of experimental data and numerical calculations \cite{CAM11,ZAC12}. Since, the $AlU$ and $NiAl$ systems share the same crystallographic f.c.c. structure,  the presented description is analogous for both alloys. The full set of frequencies are evaluated employing the echonomic Monomer method \cite{RAM06}. The Monomer \cite{RAM06} is used to compute the saddle points configurations from which we obtain the jumps frequencies defined in the 5-frequency model. Here, the inter-atomic interactions are represented by suitable EAM potentials \cite{PAS11} for the $AlU$ binary system. For the case of the $NiAl$ system, our results are in excellent agreement with the experimental data for both, the solvent self-diffusion coefficient and the solute tracer diffusion coefficient \cite{CAM11,ZAC12}. In this case we found that $Al$ in $Ni$ at diluted concentrations migrates as free species, confirmed by a weak binding between $Al$ with vacancies ($V$). Comparison of the available experimental data of the diffusion coefficient of $U$ diluted in $Al$, with the diffusion coefficient of the paired $U+V$ complex, show an excellent agreement. From theoretical evidence here presented, and from experimental data in \cite{HOU71}, we can infer that in this alloy a vacancy drag mechanism is likely to occur. Magnitudes as, the strong uranium-vacancy binding, the values of the vacancy wind at high temperatures and negative values of the cross $L$-coefficient, (give us magnitudes that)lead us to this conclusion.

The paper is organized as follows. In Section \ref{S1} we briefly introduce a summary of the macroscopic equations of atomic transport that are provided by non-equilibrium thermodynamics \cite{ALL03,HOW64}. In this way an analytical expression of the diffusion coefficients in binary alloys in terms of Onsager coefficients is presented. In section \ref{S2}, we describe the kinetic theory of isothermal diffusion process with an emphasis on the magnitudes used later. This allows to express the Onsager coefficients in terms of the frequency jumps following the procedure of Allnatt as in Ref. \cite{ALL03}. Section \ref{S3}, is devoted to give the way to evaluate the Onsager phenomenological coefficients following the procedure of Okamura and Allnat \cite{OKA83} in terms of the jumps frequencies in the context of a multi-frequency model. In Section \ref{S5}, we present expressions to evaluate the self diffusion coefficient in terms of so called solvent enhancement factor at first order in the solute concentration ($c_{S}$), and the solute diffusion coefficient is calculated at zero order in $c_{S}$. Finally, in Section \ref{S6} we present our numerical results using the theoretical procedure here summarized and showing a perfect accuracy with available experimental data, that is, for the $NiAl$ system. The last section briefly presents some conclusions.

Readers trained in this theory, can directly jump to section \ref{S5}.

\section{Theory Summary: The flux equations}
\label{S1}
Isothermal atomic diffusion in multicomponent systems
can be described by the theory of irreversible processes, in which
the main characteristic is the rate of entropy production per unit
volume $S$ \cite{ALL03}, 
\begin{equation}
TS=\sum_{k}^{N}\vec{J}_{k}.\vec{X}_{k},
\label{eq.1}
\end{equation}
 where $T$ is the absolute temperature, $N$ the number of components in the system,  $\vec{J}_{k}$ describes the flux vector density, while $\vec{X}_{k}$ is the driving force acting on component $k$. A linear expression for the flux vector $\vec{J}_{k}$ in terms of the driving forces, involves the Onsager coefficients $L_{ij}$, 
\begin{equation}
\vec{J}_{k}=\sum_{i}^{N}L_{ki}\vec{X}_{i}.
\label{eq.2}
\end{equation}
The second range tensor $L_{ij}$ is symmetric ($L_{ij}=L_{ji}$)
and depends on pressure and temperature, but is independent of the driving forces $\vec{X}_{k}$. From (\ref{eq.2}) the $1^{st}$ Fick's law, which describe the atomic jump process on a macroscopic scale, can be recovered. On the other hand, on each $k$ component, the driving forces may be expressed, in abscense of external force, in terms of the chemical potential $\mu_{k}$, so that \cite{ALL03}, 
\begin{equation}
\vec{X}_{k}=-T\nabla\left(\frac{\mu_{k}}{T}\right).\label{eq.3}
\end{equation}
Where the chemical potential $\mu_{k}$ is the partial derivative
of the Gibbs free energy with respect to the number of atoms of specie $k$,
\begin{equation}
\mu_{k}=\left(\frac{\partial G}{\partial N_{k}}\right)_{T,P,N_{j\neq k}} = \mu^{\circ}_k(T,P)+k_BT\ln(c_k\gamma_k).\label{mu}
\end{equation}
with $\gamma _k$, the activity coefficients, defined in terms of the activity 
$a_k=\gamma_{k}c_{k}$ and $c_k$ the concentration of specie $k$. For an isothermal diffusion process mediated by a vacancy mechanism, and by making use of the elimination of the dependent fluxes, 
\begin{equation}
\sum_{i}^{N}J_{i}=0\Rightarrow\sum_{k=1}^{N}L_{ki}=0.
\end{equation}
 For the particular case of a binary diluted alloy $(A,S)$ containing $N_{A}$ host atoms, $N_{S}$, solute atoms (impurities), $N_{V}$ vacancies after some algebra we arrive at the flux expressions,
\begin{eqnarray}
J_{A} & = & L_{AA}(X_{A}-X_{V})+L_{AS}(X_{S}-X_{V})\,;\label{JA}\\
J_{S} & = & L_{SA}(X_{A}-X_{V})+L_{SS}(X_{S}-X_{V})\,;\label{JB}\\
j_{V} & = & -(J_{A}+J_{S}).\label{JV}
\end{eqnarray}
Now we come back to the flux equations  (\ref{JA}-\ref{JV})
where we will introduce the chemical potential equations  (\ref{mu})
in the driving forces (\ref{eq.3}). In this way we obtain the generalized $1^{st}$ Fick's law, which includes cross effects: 
\begin{equation}
J_{A}=-\left(\frac{L_{AA}}{c_{A}}-\frac{L_{AS}}{c_{S}}\right)k_{B}T\left(1+\frac{\partial ln\gamma_{A}}{\partial lnc_{A}}\right)\nabla c_{A},
\end{equation}
 
\begin{equation}
J_{S}=-\left(\frac{L_{SS}}{c_{S}}-\frac{L_{AS}}{c_{A}}\right)k_{B}T\left(1+\frac{\partial ln\gamma_{S}}{\partial lnc_{S}}\right)\nabla c_{S},
\end{equation}
Hence, for a binary system the diffusion coefficient of the solvent and the solute $S$ are: 
\begin{equation}
D_{A}=\frac{k_{B}T}{N}\left(\frac{L_{AA}}{c_{A}}-\frac{L_{AS}}{c_{S}}\right)\phi_{A} = D^{\star}_A\phi_{A} ,
\label{DATEQ}
\end{equation}
\begin{equation}
D_{S}=\frac{k_{B}T}{N}\left(\frac{L_{SS}}{c_{S}}-\frac{L_{SA}}{c_{A}}\right)\phi_{S} = D^{\star}_S\phi_{S} .\label{DBTEQ}
\end{equation}
and 
\begin{equation}
D_{V}=\frac{k_{B}T}{c_{V}}\left(L_{AA}+L_{SS}+2L_{AS}\right).\label{DVTEQ}
\end{equation}.
$D_{A},D_{S}$ are commonly known as the intrinsic diffusion coefficients, while $D^{\star}_{A}$ and $D^{\star}_{S}$ are the isotopic tracers diffusion coefficients that are the magnitudes experimentally measured. $D_{S}$ is the vacancy diffusion coefficient. In the spite of Gibbs-Duhem relation, 
\begin{equation}
\sum _k N_kX_k=0,
\end{equation}
the thermodynamic factors $\phi_A,\, \phi_S$ are equal:
\begin{equation}
\phi_{A}=\left(1+\frac{\partial ln\gamma_{A}}{\partial lnc_{A}}\right)=\phi_{S}=\left(1+\frac{\partial ln\gamma_{S}}{\partial lnc_{S}}\right)=\phi_{0}. 
\end{equation}
\noindent We are interested in diluted alloys, that is, in the
limit $c_{S}\rightarrow0$ where $\phi_{0}=1$.  The solute diffusion
coefficient is calculated directly from the intrinsic one through
the expression, 
\begin{equation}
D^{\star}_S=D_{S}=\frac{1}{c_{S}}\left(\frac{k_{B}T}{N}L_{SS}\right)\,\,\,;\,\,\, c_{S}\rightarrow0,\label{DB2TEQ}
\end{equation}
while for the solvent $D^{\star}_{A}$, is calculated from (\ref{DATEQ}).

In the next sections, we express these last Onsager coefficients in terms of microscopical atomic jump frequencies. 

\section{The Kinetic Equations}
\label{S2} 

In this section we present a brief description
of the applicability of the master equation to atomic transport
in metals in terms of the spatial distribution of atoms and defects \cite{ALL03}.
The theory provides specific results to evaluate the atomic Onsager
transport coefficients for systems in which there is an attractive
interaction between solute and vacancies. The solute-vacancy pair
is identified by the subscripts $p,q$. Where $p,q$ denotes the sites
in the lattice where the solute and vacancy are locate respectivelly.
By configuration we mean any distinct orientation of the pair .

We suppose that the solute-vacancy defect changes from $p$ to $q$
by thermal activation at a rate $\omega_{qp}$. These transition are
taken to be Markovian, i.e, the $\omega_{qp}$ depend on the initial
and final configurations but are independent of all previous transitions.
We denote the number density of defects which are in configuration
$p$ at time $t$ by $n_{p}(t)$. For a closed set of configuration
$p$ the rate equations for the densities $n_{p}(t)$ are, 
\begin{equation}
\frac{\partial n_{p}(x,t)}{\partial t}=-\sum_{q\neq p}\omega_{qp}n_{p}+\sum_{q\neq p}\omega_{pq}n_{q}.
\label{MatP}
\end{equation}
 The first term represents the rate at which the vacancy in $p$ leave
the site to all the other configurations $q$ ($q\neq p$) first neighbors
of $p$. The second is the rate at which the vacancy reaches $p$
from $q$. Here $n_{p}(t)=n_{d}p_{p}(t)$ where $n_{d}$ is the defect
density independently of its configuration and $p_{p}(t)$ is defined as the fraction of all defects that are in configuration
$p$ at time $t$. In matrix notation equation (\ref{MatP})
is, 
\begin{equation}
\frac{dp_{p}(x,t)}{dt}=-Pp_{p}(t),
\label{eq:fp}
\end{equation}
 where $p_p$ is a column matrix whose elements are the probabilities
of occupation, $p_p\equiv\left\{ p_{1},p_{2},...\right\} $, while $P$
is defined as follows: 
\begin{eqnarray}
P_{qp} & = & -\omega_{qp}\,;\,(q\neq p)\\
P_{pp} & = & \sum_{q\neq p}\omega_{qp}.
\label{Pcoef}
\end{eqnarray}
One feature of this equation, that we will be used later, is that we can solve equation 
(\ref{MatP}) in terms of a reduced matrix $Q$, which can be obtained from $P$ such that its matrix elements $Q_{pq}$ are 
\begin{equation} 
Q_{pq}=P_{pq} -P_{p\overline{q}},
\end{equation} 
now the indexes $p$ and $q$ only take positive values, that is, jumps that involve a drift in the positively defined sense of the principal crystal axis, while jumps in the opposite direction are denoted by  $overline{q}$. In this way, $Q$ is a $n\times n$ dimension square matrix, where $n$ is the number of different configurations in the positive principal crystal axis minus 1, as we will see in next section. Under thermal equilibrium $P\equiv P^{(0)}$, is given by statistical thermodynamics as, 
\begin{equation}
p_{p}^{(0)}=\frac{\exp(-E_{p}^{(0)}/k_{B}T)}{\sum_{\gamma}\exp(-E_{\gamma}^{(0)})/k_{B}T}\,;\,(\forall p)
\end{equation}
in which $E_{\gamma}^{(0)}$ is the Gibbs energy of the system in state $\gamma$. Under the same conditions we write in the
steady state ($\frac{dp_{p}(x,t)}{dt}=0$), the principle of detailed
balance which gives us the useful relation, 
\begin{equation}
\omega_{qp}^{(0)}p_{p}^{(0)}=\omega_{pq}^{(0)}p_{q}^{(0)}\,;\,(\forall p,q)
\label{DB}
\end{equation}
 that is,
\begin{equation}
\frac{\omega_{qp}^{(0)}}{\omega_{pq}^{(0)}}=\exp(E_{p}^{(0)}-E_{q}^{(0)}).
\end{equation}
Where supraindex $(0)$ denote magnitudes in the thermodynamical equilibrium.

From the master equations, it is possible to calculate mean values and second momenta of the basic kinetic quantities, (ex. $x$ coordinate of a tracer atom) in the thermodynamic equilibrium. Also it is very useful to perform averages in regions that are small in a macroscopic sense, but large enough to contain many lattice points.
Then, solving the master equation in the linear response approximation it is possible to obtain formal expression for the transport coefficients and to verify the Onsager relations.  In this way,  the macroscopic flux equations and the transport coefficients may be expressed in terms of averaged microscopic variables.  This is indeed a generalization of the Einstein relations for the Brownian motion. Also, it permits to express the Onsager coefficients in terms of the jump frequencies. This procedure is described in details in \cite{ALL03}.

It is now very useful to introduce the expressions derived by Franklin and and Lidiard \cite{FRA83} for the Onsager coefficients and kinetic theory, in terms of the reduced $Q$ matrix. The authors wrote equations for the fluxes $J_S$ and $J_D$ ($D$ can be vacancies, $V$, or interstitials, $I$) in terms of thermodynamical forces, which are precisely of the form required by non-equilibrium thermodynamics, then up to second-virial coefficients. The, let the Onsager coefficient obtained defined as, 
\begin{equation}
\begin{array}{ccl}
\frac{k_BT}{N}L_{KM} & = &\frac{1}{2}\sum _{p,q} a_{qp}^K a_{qp}^M \omega _{qp}p^{(0)}_p +\sum _{d,p} a_{dp}^Ka_{dp}^M\omega _{dp}p^{(0)}_p \vspace{0.5cm}\\
& + & \frac{1}{2}\delta_{KM}p^{(0)}_K(1-z_{f}p^{(0)}_{\overline{K}})\sum_r(a^K_{r0}))^2\omega^{K(0)}_{r0}\vspace{0.5cm} \\
& - &2\sum_{p,q}^{(+)}v_q^K(Q^{-1})_{qp}v_p^Mp^{(0)}_p.
\end{array}
\label{eq:Lij}
\end{equation}
The velocity function is defined as,
\begin{equation}
v_p^{M}=\sum _{u} a_{up}^M\omega _{up} + \sum _{p} a_{dp}^M\omega _{dp}.
\label{eq:vij}
\end{equation}

The subscripts $K,M$ each of which may be either $A$, $S$ or $D$, where $A,S$ represent the solvent or solute atoms, while $D$ stand for the defects that may be either vacancies or interstitials. We use the same labeling $p$,$q$ and $u$ for the paired species as before,  and $r$ for unpaired species that can be of type $S$ (free solute) or $D$ (free defects). While the label $d$ (second term) takes into account  dissociative jumps, that is, it runs on sites that after the jump the defect is unpaired with the solute. The assumption at which the species are regarded to be paired or free may be fixed arbitrarily. The jump distance in a $p \rightarrow q$ transition  are represented by $a^{K,M}_{qp}$, they take account the movement of the both, the $K$ and $M$ species. Similarly, $a^{K(0)}_{r0}$ are the jump distance of the free species, and $a^{K,M}_{dp}$, stand for the distance of dissociative jumps.

The first term on the right side in (\ref{eq:Lij}), is the uncorrelated contribution of transitions from one paired configuration to another. The second term gives the sum of the two corresponding contributions from dissociation and association transition (equal by detailed balance, hence no factor $1/2$ as in the first term), while the third term is the uncorrelated contribution from the free-free transition (corrected by the term in $z_f$ for the fact that some movements of the unassociated pair may result in the formation of an associated pair, the contribution of which have already been accounted for in the second term). 

Correlated movements are represented in the fourth term, which contain the $Q$ matrix. $\overline{K}$ means not $K$ (i.e., if $\overline{K}=V$, then $K=S$ and vice versa).  Note that the summation only runs over those pair configurations in the $(+)$ set, that is, jumps that involve a drift in the positively defined sense of the principal crystal axis.  Although the summations contained in the velocity analogue $v^M_p$ are over all configurations which can be reached in one transition from a configuration $p$ lying in the $(+)$ set. We note that the relevant point is to obtain the reduced $Q$ matrix from Kinetic theory.

Below we apply the formalism following the procedure described by
Allnat and Lidiard \cite{ALL03} to calculate the Onsager coefficients $L_{AA}$,
$L_{SS}$ and $L_{AS}=L_{SA}$ and therefore the tracer diffusion
coefficients $D_{A}^{\star}$ and $D_{S}^{\star}$.

\section{The $L$-coefficients in the shell approximation}
\label{S3} 
\begin{figure}[h]
\begin{center}
\includegraphics[width=8.0cm]{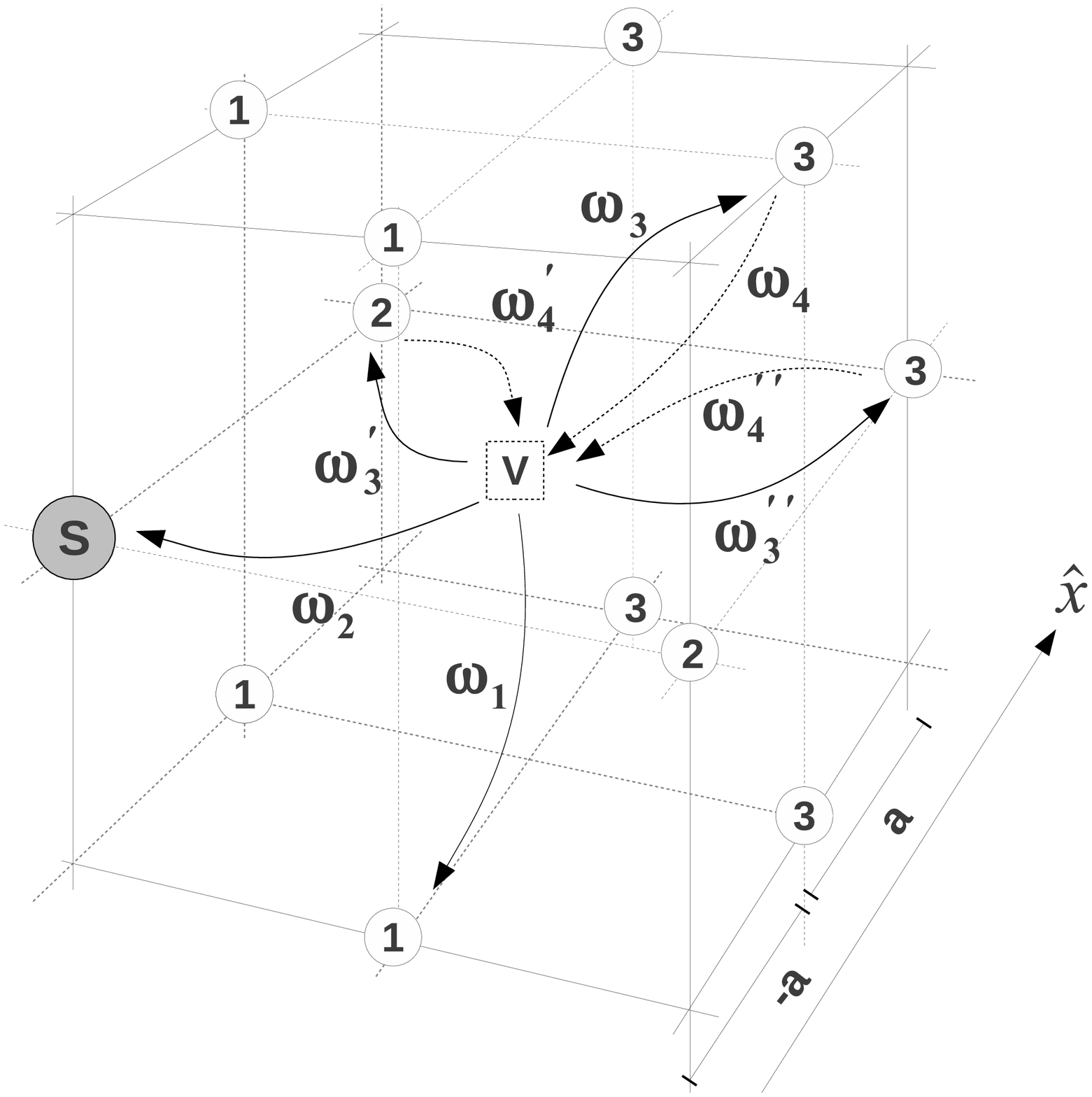}
\vspace{-1.0cm}
\caption{The five-frequency model of a solute-vacancy pair in a f.c.c. lattice.}
\label{FIG1}
\end{center}
\end{figure}

\begin{figure}[h]
\hspace{-1.cm}\includegraphics[width=14.0cm]{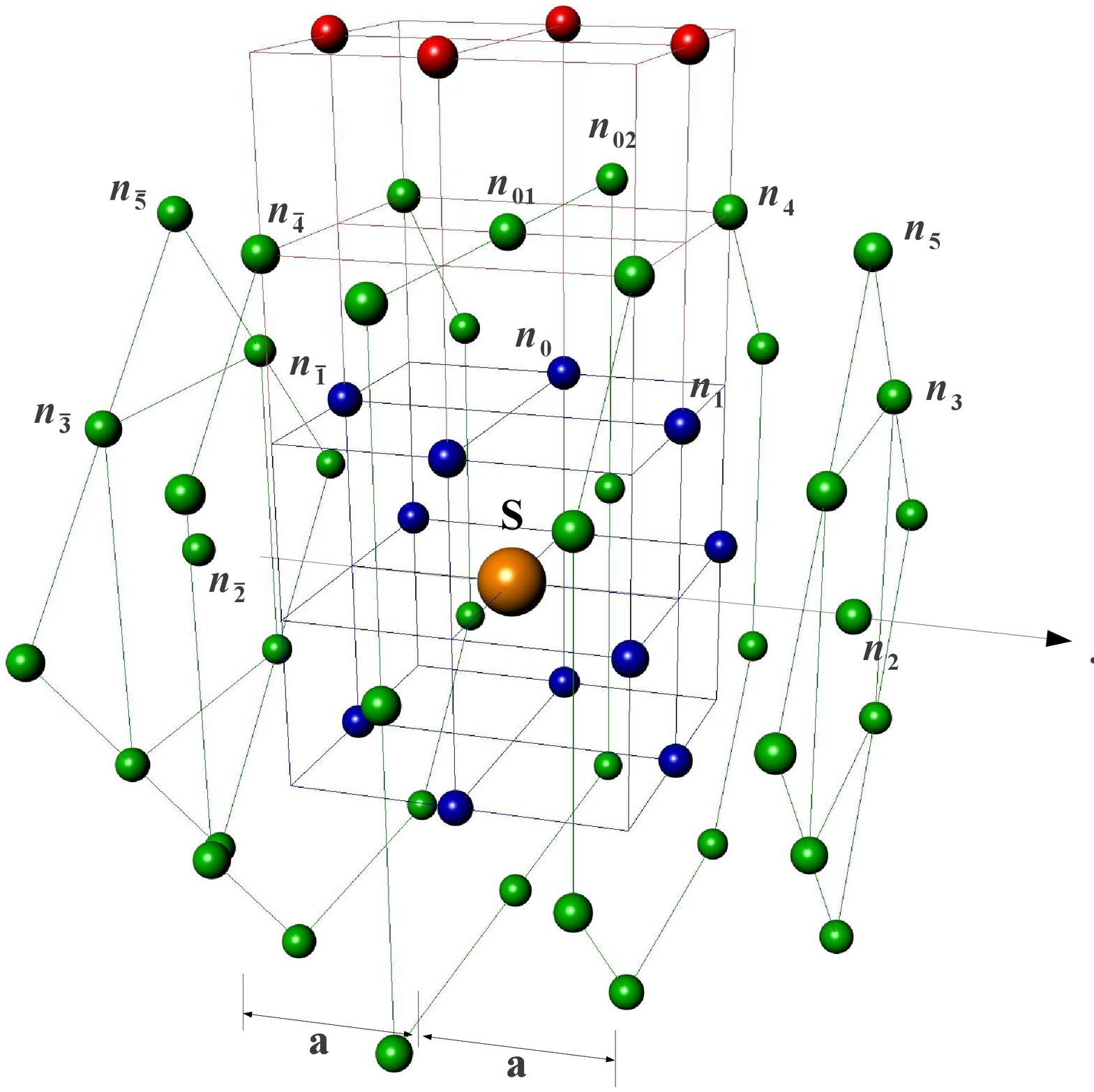}
\caption{The coordinated shell model in f.c.c. lattice (see Ref. \cite{BOC74}). The different types of symmetries shown are detailed in Table \ref{T2}. In the figure, blue bullets are the first twelve neighbors sites to the solute $\bf{S}$ at the origin. In green the 42 subsequent sites. In red, the third coordinated shell from which the vacancy never returns to the second shell. }
\label{FIG2}
\end{figure}

Here we assume that the perturbation of the solute movement by a vacancy $V$, is limited to its immediate vicinity, hence we adopt an effective five frequency model \`a la Le Claire \cite{LEC78} for f.c.c. lattices, to understand the effect of different vacancy exchange
mechanisms on solute diffusion. In such a model, the frequencies jumps
$\omega_{qp}$ are now denoted only with one index $\omega_{i}$ ($i=0,1,2,3,4$). In Figure \ref{FIG1} the jump rates are indicated as $\omega_{i}$
($i=1,2,3,4$). We suppose that all gradient potential and concentration are along a particular crystal principal axis $\hat{x}$. 
Considering only jumps between first neighbors, for them, $w_{2}$ implies in the exchange between the vacancy and the solute, $w_{1}$ when the exchange between the vacancy and the solvent atom lets the vacancy as a first neighbor to the solute (positions denoted with circled 1 in figure \ref{FIG1}). The frequency of jumps such that the vacancy goes to sites that are second neighbor of the solute is denoted by $\omega_{3}$ (sites with circled 2). The model includes the jump rate $\omega_{4}$ for the inverse of $\omega_{3}$. Jumps toward sites that are third and forth neighbor of the solute are denoted by $\omega_{3}^{\prime}$ and $\omega_{3}^{\prime\prime}$ respectively while $\omega_{4}^{\prime}$ and $\omega_{4}^{\prime\prime}$ are used for their respective inverse frequency jumps. The jump rate $\omega_{0}$ is used for vacancy jumps among sites more distant than forth neighbors of the solute atom. In this context, that enables association ($\omega_{4}$) and dissociation
reactions ($\omega_{3}$), i.e the formation and break-up of pairs,
the model include free solute and vacancies to the population of bounded
pairs. It is assumed that a vacancy which jumps from the second to the third shell, with $\omega_0$, never returns (or does so from a random direction). As in Ref. \cite{CHO11} we express 
\begin{equation}
\omega^{\star}_3=2\omega_3 + 4\omega^{\prime}_3 + \omega^{\prime\prime}_3,
\label{w3eff}
\end{equation}
 and 
\begin{equation}
\omega^{\star}_4=2\omega_4 + 4\omega^{\prime}_4 + \omega^{\prime\prime}_4.
\label{w4eff}
\end{equation} 
This procedure allows us to usufruct Boquet equations \cite{BOC74}, and the technique develloped by Allnatt and Lidiard to evaluate the transport coefficients of dilute solid solutions \cite{ALL03}. The six symmetry types of vacancy sites that are in the first coordination shell (first neighbor of the solute) or the second coordination shell (sites accessible from the first shell by one single vacancy jump) are listed in Table \ref{T1} (using the same notation as in Ref. \cite{OKA83}) and plotted in Figure \ref{FIG2}.
\begin{table}[h]
\begin{center}
\caption{Symmetry types for f.c.c. lattice for vacancy-sites at the first
4 nearest$-$neighbor separation from the impurity $S$ at the origin
(Ref. \cite{OKA83}). The forth and fifth columns denote the velocity
functions of the solvent $v_{p}^{(A)}$ and solute atoms $v_{p}^{(B)}$respectively
devided by the spacing parameter $a$ (see text below).}
\label{T1}
\begin{tabular}{c|lccc}
\hline 
\, Symmetry type $i$ \, & \, vacancy-sites (Ref. \cite{BOC74}) \, &  \, n.n.s. \, & $v^{(A)}_p/a$ & \, $v^{(S)}_p/a$ \,  
\tabularnewline
\hline 
\hline 
\, 1  \, & \, $(1,1,0)$,$(1,\overline{1},0)$,$(1,0,1)$,$(1,0,\overline{1})$ \, & \, 1 \, & $(2\omega_1-3\omega^{\star}_3)$ & \, $\omega_2$ \,
\tabularnewline 
\, 2 \, & \, $(2,0,0)$ \, & \, 2 \, & $4(\omega^{\star}_4-\omega_0)$ & \, $0$ \,
\tabularnewline 
\, 3 \, & \, $(2,1,1)$,$(2,1,\overline{1})$,$(2,\overline{1},1)$,$(2,\overline{1},\overline{1})$ \, & \, 3 \, & $2(\omega^{\star}_4-\omega_0)$ & \, $0$ \,
\tabularnewline 
\, 4 \, & \, $(1,2,1)$,$(1,2,\overline{1})$,$(1,\overline{2},1)$,$(1,\overline{2},\overline{1})$ \, & \, 3 \,  & $(\omega^{\star}_4-\omega_0)$ & \, $0$ \,
\tabularnewline 
\, \,   & \, $(1,1,2)$,$(1,1,\overline{2})$,$(1,\overline{1},2)$,$(1,\overline{1},\overline{2})$ \, & \,  \, 
\tabularnewline 
\, 5 \, & \, $(2,2,0)$,$(2,\overline{2},0)$,$(2,0,0)$,$(2,0,\overline{2})$ \, & \, 4 \, & $(\omega^{\star}_4-\omega_0)$ & \, $0$ \,
\tabularnewline
\hline 
\end{tabular}
\end{center}
\end{table}
As usual \cite{BOC74}, sites that are equally distant from the solute atom $S$ at the origin, and that have the same abscissa (x-coordinate in Fig.\ref{FIG2}) share the same vacancy occupation probability $n_i$, $n_{\overline{i}}$. Table \ref{T2} resumes the here employed notation. Here, we denote the sites probability with $n_{ij}$ where for $i\neq 0$ there is only one index $i$ that is given in crescent order in the distance to the solute atom $S$. Also, non overlined indexes imply in a positive abscissa, while overlined ones $\overline{i}$ denote sites with negative $x$ coordinate. For the sites in the $x=0$ plane ($i=0$), the sites are denoted with two subindexes $n_{0j}$, where the second index $j$ is given in crescent order of the distance to the solute atom $S$. Table \ref{T2} denotes the number of different types of sites and the distance of them to the $x$ axis.
\begin{table}[h]
\begin{center}
\caption{Probability of occurrence of the vacancy at a site of the subset $n_j$.}
\label{T2}
\begin{tabular}{c|ccccccccccccc}
\hline 
\, $n_{ij}$ (Ref. \cite{BOC74}) \, & \, $n_{5}$ \, &  \, $n_{4} \,$ &  \, $n_{3}$ \, &  \, $n_{2}$ \, &  \, $n_{1}$ \,  &  \, $n_{01}$ \,  &  \, $n_{02}$ \, &  \, $n_{0}$ \, &  \, $n_{\overline{1}}$ \,&  \, $n_{\overline{2}}$ \,&  \, $n_{\overline{3}}$ \,&  \, $n_{\overline{4}}$ \,&  \, $n_{\overline{5}}$ \, \tabularnewline
\hline 
\hline 
\, $\#$ of sites \, & \, 4 \, &  \, 8 \, &  \, 4 \, &  \, 1 \,  &  \, 4\,  &  \, 4 \, &  \, 4 \, & \, 4 \, & \, 4 \,& \, 1 \,& \, 4 \,& \, 8 \,& \, 4 \,  
\tabularnewline
\hline 
\, separation \, & \, $2a$ \, &  \, $a\sqrt{2}$ \, &  \, 0 \, &  \, $a\sqrt{5}$ \,  &  \, $a$\,  &  \, $a\sqrt{2}$ \, &  \, $2a$ \, & \, $a\sqrt{2}$ \, & \, $a$ \,& \, $a\sqrt{5}$ \,& \, 0 \,& \, $a\sqrt{2}$ \,& \, $2a$ \,  
\tabularnewline
\hline 
\end{tabular}
\end{center}
\end{table}
With this classification, the basic kinetic equations (\ref{MatP}) for the first coordinated shell approximation \cite{OKA83} in the steady state are written as,
\begin{eqnarray}
\frac{\partial n_1}{\partial t} & = & -(2\omega_1+\omega_2+7\omega^{\star}_3)n_{1} + \omega^{\star}_4n_{2} + 2\omega^{\star}_4n_{3} + 2\omega^{\star}_4n_{4} + \omega^{\star}_4n_{5} + 2\omega_1n_{0} + \omega_4n_{01} + \omega_2n_{\overline{1}}=0, \, \nonumber \\
\frac{\partial n_{2}}{\partial t} & = & 4\omega^{\star}_3n_{1} -(8\omega_0+4\omega^{\star}_4)n_{2}+ 4\omega_0n_{3}=0, \, \nonumber \\
\frac{\partial n_{3}}{\partial t} & = & 2\omega^{\star}_3n_{1} + \omega_0n_{2} -(10\omega_0+2\omega^{\star}_4)n_{3}+ 2\omega_0n_{4} + \omega_0n_5=0, \, \nonumber  \\
\frac{\partial n_{4}}{\partial t} & = & \omega^{\star}_3n_{1} +\omega_0n_{3} -(9\omega_0+2\omega^{\star}_4)n_{4}+ \omega_0n_{5}+\omega^{\star}_3n_{0} + \omega_0n_{01} + \omega_0n_{02}=0, \, \nonumber   \\
\frac{\partial n_{5}}{\partial t} & = & \omega^{\star}_3n_{1} +4\omega_0n_{3} +2\omega_0n_{4} -(8\omega_0+\omega^{\star}_4)n_{5}=0, \,\label{set0} \\
\frac{\partial n_{0}}{\partial t} & = & 2\omega_1n_{1} +2\omega^{\star}_4n_{4}-(4\omega_1+7\omega^{\star}_3)n_{0} + 2\omega^{\star}_4n_{01} + \omega^{\star}_4n_{02} + 2\omega_1n_{\overline{1}} +2\omega^{\star}_4n_{\overline{4}}=0, \, \nonumber  \\
\frac{\partial n_{01}}{\partial t} & = & \omega_1n_{1} +2\omega_0n_{4} +2\omega_0n_{0} -(8\omega_0+4\omega^{\star}_4)n_{01}+ \omega_3n_{\overline{1}}+2\omega_0n_{\overline{4}}=0 , \, \nonumber  \\
\frac{\partial n_{02}}{\partial t} & = & 2\omega_0n_{4} +\omega_3n_{0} +\omega_0n_{01} -(11\omega_0+\omega^{\star}_4)n_{02} + 2\omega_0n_{\overline{4}}=0, 
\nonumber  \\
 & \vdots &  \nonumber
\end{eqnarray}
\noindent where the vertical dots denotes the analogous sets of equations for the overlined indexes $\partial n_{\overline{i}} / \partial t =0$. Hence, the matrix $P$ defined in (\ref{Pcoef}) that stands from equation (\ref{set0}) is such that  $P \in R ^{13 \times 13}$. Then the reduced matrix $Q$, whose elements are $Q_{pq}$, can be obtained from $P$ as, 
\begin{equation} 
Q_{pq}=P_{pq} -P_{p\overline{q}},
\end{equation} 
now the indexes $p$ and $q$ take only positive values, such that $Q$ is a five dimension square matrix, so that 
\begin{eqnarray}
Q & = & \left (\begin{array}{ccccc}
(2\omega_1 + 2\omega_2 + 7\omega^{\star}_3) & -4\omega^{\star}_4 & 2\omega^{\star}_4 & - \omega^{\star}_4 & -\omega^{\star}_4 \\
- \omega^{\star}_3 & (4\omega^{\star}_4 + \omega_0) & - \omega_0 & 0 & 0\\
- 2\omega^{\star}_3 & - 4\omega_0 & (2\omega^{\star}_4 + \omega_0) & -\omega_0 & -\omega_0 \\
- 2\omega^{\star}_3 & 0 & - 2\omega_0 &  (2\omega^{\star}_4 + \omega_0) & -2\omega_0 \\
- \omega^{\star}_3 & 0 & - 2\omega_0 & -\omega_0 & (\omega^{\star}_4 + \omega_0) 
\end{array} \right )
\label{Q}
\end{eqnarray}
\noindent The matrix element $(Q^{-1})_{11}$ of the inverse matrix of $Q$ defines a factor named $F$, introduced by Manning \cite{MAN68}, 
\begin{equation} 
(Q^{-1})_{11}=(2\omega_1+2\omega_2+7\omega^{\star}_3F)^{-1}. 
\label{F}
\end{equation}
\noindent The quantity $1-F$ is the fractional reduction in the overall frequency of jumps from a first-shell site to a second-shell site caused by returns of vacancy to first-shell sites,
\begin{equation}
7(1-F)=\frac{10\epsilon^{4}+ B_1\epsilon^{3}+B_2\epsilon^{2}+B_3\epsilon}{2\epsilon^{4}+ B_4\epsilon^{3}+B_5\epsilon^{2}+B_6\epsilon +B_7}
\label{FW}
\end{equation}
where $\epsilon = \omega^{\star}_4/\omega_0$ and Table \ref{T3} shows the $B_i$ coefficients calculated by Koiwa \cite{KOI83,ALL03} and that will be employed in the present calculations. 
\begin{table}[h]
\begin{center}
\caption{Coefficients in the expression for $F$ for the five \\
frequency model calculated by Koiwa \cite{KOI83}.}
\label{T3}
\begin{tabular}{c|ccccccc}
\hline 
\, \, & \, $B_1$ \, & \, $B_2$ \, & \, $B_3$ \, & \, $B_4$ \, & \, $B_5$ \, & \, $B_6$ \,& \, $B_7$ \,  
\tabularnewline
\hline 
\hline 
\,  Ref. \cite{KOI83} \, & \, 180.3 \, & \, 924.3 \, & \, 1338.1 \, & \, 40.1 \, & \, 253.3 \, & \, 596.0 \, & \, 435.3 \,
\tabularnewline 
\hline 
\end{tabular}
\end{center}
\end{table}
\newline
For evaluating the $L$-coefficients, we shall need both the site fraction $c_p$ of solute atoms which a vacancy among their $z$ nearest-neighbor sites, also the fraction of unbounded vacancies $c^{\prime}_V=c_V-c_p$ and of unbound solute atoms $c^{\prime}_S=c_S-c_p$. These are related through the mass action equation, namely
\begin{equation}
\frac{c_p}{c^{\prime}_Vc^{\prime}_S} = z \exp(-E_b/k_BT)=\frac{\omega^{\star}_4}{\omega^{\star}_3}.
\label{cpg}
\end{equation}
With $E_b$ the binding energy of the solute  vacancy  pair related to $\omega^{\star}_4/\omega^{\star}_3$ by the use of detailed balance. Then, if the pairs and free vacancies are in local equilibrium and, since the fraction of solute $c_S$ will be much greater than $c_V$ and thus also $c_p$, we can express the equilibrium constant $K$ as,
\begin{equation}
\frac{c_p}{c_V-c_p}=zc_S\exp(-E_b/k_BT)\equiv Kc_S,
\end{equation}
and equivalently
\begin{equation}
c_p=c_V\left( \frac{Kc_S}{1+Kc_S}\right).
\label{cp0}
\end{equation}
The Onsager coefficients can be entirely obtained from equation (\ref{eq:Lij}) in terms of the concentration of free and paired species, and in terms of the jump frequency rates $\omega _i$.
For the case of binary alloys in f.c.c. lattices, symmetry arguments and spacial isotropy implies that the only needed coefficients are $L_{AA}$, $L_{SS}$ and $L_{AS}$. In this respect, the velocity terms are depicted in the forth and fifth column  of Table \ref{T1}. For the case where the Onsager coefficients are expressed in terms of the five-frequency model, the only required elements of $Q^{-1}$  is  $(Q^{-1})_{11}$ all the other elements appearing in $L_{AS}$ and $L_{AA}$ can be eliminated \cite{ALL03}. Hence, the Onsager coefficients (\ref{eq:Lij}) are \cite{OKA83}
\begin{equation}
L_{AA}=\frac{Ns^2}{6k_BT}\left\{ 12c^{\prime}_V (1-7c^{\prime}_S)\omega_0+c_p(A^{(0)}_{AA}+A^{(1)}_{AA})\right\}
\label{LAA1F}
\end{equation}
\begin{equation}
L_{AS}=L_{SA}=c_pA^{(1)}_{AS}
\label{LAB1F}
\end{equation}
\begin{equation}
L_{SS}=\frac{Ns^2c_p\omega_2}{6k_BT}\left\{1-\frac{2\omega_2}{\Omega}\right\}
\label{LBB1F}
\end{equation}
where $s=a_A/\sqrt{2}$ is the jump distance, $a_A$ the lattice parameter of solvent $A$. The concentration of free solute and vacancy defects are denoted with $c^{\prime}_S$ and $c^{\prime}_V$ respectively. While, 
\begin{equation}
\Omega=2(\omega_1+\omega_2)+7\omega^{\star}_3F.
\end{equation}
We define the solute correlation factor $f_S$ for the bracket in (\ref{LBB1F}) as,
\begin{equation}
f_S= 1-\frac{2\omega_2}{2(\omega_1+\omega_2)+7\omega^{\star}_3F}.
\label{eq:FF}
\end{equation}
\noindent Completing the definitions in (\ref{LAA1F}) and (\ref{LAB1F}) with,
\begin{equation}
A^{(0)}_{AA}=4\omega_1+14\omega^{\star}_3 
\end{equation}
\begin{eqnarray}
A^{(1)}_{AA} & = & \frac{1}{\Omega} \left[ -2(3\omega^{\star}_3-2\omega_1)^2+14\omega^{\star}_3(1-F) \right. \left(\frac{\omega_0-\omega^{\star}_4}{\omega_4}\right) \nonumber \\
           & \times & \left. \left\{(3\omega^{\star}_3-2\omega_1) - 2(\omega_1+\omega_2+7\omega^{\star}_3/2)\left(\frac{\omega_0-\omega^{\star}_4}{\omega^{\star}_4}\right)\right\} \right] 
\end{eqnarray}
\begin{equation}
A^{(1)}_{AS}=\frac{\omega_2}{\Omega}\left[ 2(3\omega^{\star}_3-2\omega_1)+14\omega^{\star}_3(1-F)\left(\frac{\omega_0-\omega^{\star}_4}{\omega^{\star}_4}\right) \right].
\end{equation}

In order to calculate the self diffusion coefficients $D^{\star}_A$ and $D^{\star}_S$ we must replace the $L$-coefficients expressions (\ref{LAA1F},\ref{LAB1F},\ref{LBB1F}) in (\ref{DATEQ},\ref{DBTEQ}), that for the diffusion coefficients. 

\section{Expressions for $D^{\star}_{A}$, $D^{\star}_{S}$ and $D^{\star}_{p}$ coefficients}
\label{S5}
A comparison between experimental data and the present simulations are possible with the knowledge of the two tracer diffusion coefficients $D^{\star}_{A}$ and $D^{\star}_S$. For $D^{\star}_{A}$ or equivalently $L_{AA}$ it is necessary to consider the motion of the tracer atom $A^{\star}$ via a vacancy mechanism caused by both, vacancies at first neighbors of $S$ or at the unperturbed lattice sites. The tracer self-diffusion coefficient $D^{\star}_A(c_S)$ of the specie $A$ in a diluted alloy with a concentration $c_S$ of solute atoms $S$, can be written in terms of the self diffusion coefficient $D^{\star}_A(0)$, of the specie $A$ in pure f.c.c. lattice as,
\begin{equation}
D^{\star}_{A}(c_S)=D^{\star}_{A}(0)(1+b_{{A}^{\star}}c_S),
\label{DAenh}
\end{equation}
at first order in $c_S$. The solvent enhancement factor, $b_{A}$, is obtained in terms of the properties of the solute-vacancy model. On the other hand, for the pure solution, the self diffusion coefficient $D^{\star}_{A}(0)$ is given by \cite{LEC78},
\begin{equation}
D^{\star}_{A}(0)=a_A^2c^{0}_Vf_0\omega_0.
\end{equation}
where $a_A$ is the solvent lattice parameter, $f_0=0.7815$ is the correlation factor for the self-diffusion in f.c.c. lattices, and $c^0_V$ is the vacancy concentration at the thermodynamical equilibrium. This former is such that,
\begin{equation}
c^{0}_V=\exp\left( -\frac{ E^V_f}{k_BT}\right),
\label{cv0}
\end{equation}
where $T$ is the absolute temperature, $E^V_f$ is the formation energy of the vacancy in pure $A$. The entropy terms are set to zero, which is a simplifying approximation.
So that, inserting (\ref{cv0}) we get 
\begin{equation}
D^{\star}_A(c_S)=a_A^2f_0\omega_0\exp\left( -\frac{ E^V_f}{k_BT}\right).
\label{DACB}
\end{equation}
\noindent We assume $c_S \rightarrow 0$ then, we use pure lattice parameters for all our calculations. The solute-enhancement factor $b_{A^{\star}}$, is obtained by replacing (\ref{LAA1F}) in (\ref{DATEQ}) up to first order in the solute concentration. In the particular case of the five-frequency model, the expressions for the Onsager coefficients are  (\ref{LAA1F},\ref{LAB1F},\ref{LBB1F}). Hence, as in Ref. \cite{ALL03,LEC78},  we get,
\begin{eqnarray}
b_{A^{\star}} & = & -19+\frac{4\omega_1+14\omega^{\star} _3}{\omega_0}\left(\frac{\omega^{\star}_4}{\omega^{\star}_3}\right) +\frac{1}{(\omega_1+\omega_2+7\omega^{\star}_3F/2)}  \left\{ \right. \nonumber \\ 
&- & \frac{14(1-F)(\omega_0-\omega^{\star}_4)^2\times(\omega_1+\omega_2+7\omega^{\star} _3/2)}{\omega_0\omega^{\star}_4 } \label{eq:bA}\\
&+& \left. \frac{\omega^{\star}_4 (3\omega^{\star}_3-2\omega_1)^2+14\omega^{\star}_3(1-F)\times(3\omega^{\star}_3-2\omega_1)(\omega_0-\omega^{\star}_4)}{\omega_0\omega^{\star}_3}\right\} . \nonumber
\end{eqnarray}
In the diluted limit ($c_S \rightarrow 0$), $D^{\star}_S$ is identical to the intrinsic diffusion coefficient $D_S$ given by (\ref{DB2TEQ})
\begin{equation}
D_S=D^{\star}_S=\frac{k_BT}{n_S}L_{SS}.
\label{dif}
\end{equation}
Introducing $L_{SS}$ in (\ref{LBB1F}) and the detailed balance equation (\ref{cpg}) in (\ref{dif}), we obtain an expression for the tracer solute diffusion coefficient,
\begin{equation} 
D^{\star}_S=a^2\omega_2\left(\frac{c_p}{3c_S}\right)\times \left\{\frac{\omega_1+7\omega_3F/2}{\omega_1+\omega_2+7\omega_3F/2}\right\} \, = \, a^2\omega_2\left(\frac{c_p}{3c_S}\right)\times f_S.
\label{DBB2}
\end{equation}
In (\ref{DBB2}) we introduce the solute correlation factor $f_S$ as, 
\begin{equation}
f_S= \left\{\frac{\omega_1+7\omega_3F/2}{\omega_1+\omega_2+7\omega_3F/2 }\right\}.
\label{eq:FF1}
\end{equation}
where $F$ was previously defined in (\ref{F}). In the Le Claire description, $D^{\star}_S$ can also be expressed as,
\begin{equation}
D^{\star}_S=a_A^2f_S\omega_2 \exp \left( -\frac{E^V_f+E_b}{k_BT}\right).
\label{DS22}
\end{equation}
For the drift of solutes in a vacancy flux we shall make contact with the alternative
phenomenology offered by Johnson and Lam \cite{JOH76}. In terms of thermodynamic forces, which are precisely of the form required by non-equilibrium thermodynamics, up to second-virial coefficients, the flux of solute atoms $J_B$ is expressed as
\begin{equation}
J_S=-D_{p}\nabla C_{p} + \sigma_V c^{\prime}_S D_V \nabla c^{\prime}_V,
\label{JP}
\end{equation}
The coefficients $D_p$ and $D_V$ are interpreted as diffusion coefficients of pairs and free vacancies, respectively, while $\sigma_V$ is a sort of cross section for vacancies to induce solute motion. When we insert the appropriate chemical potential gradients (see Franklin and Lidiard \cite{FRA84}) into the thermodynamic flux equation (\ref{JB}), we find that (\ref{JB}) is equivalent to (\ref{JP}) if
\begin{equation}
D_{p}=\frac{k_BT}{Nc_p}L_{SS}
\label{Dp}
\end{equation} 
and
\begin{equation}
\sigma_VD_V=\frac{k_BT}{N}\frac{L_{AS}+2L_{SS}}{c_p}\left(\frac{12\omega ^{\star}_4}{\omega ^{\star}_3}\right).
\label{SIGV}
\end{equation} 
We see that for a vacancy mechanism, solute atoms may only move when they are paired with a vacancy and it is reasonable therefore that $D_S$ should be equal to 
($c_p/c_S$) as (\ref{DBTEQ}) and (\ref{Dp}) require. To obtain $\sigma _V$ from (\ref{SIGV}), we need the full expressions for $L_{AA}$ and $L_{AS}$ in (\ref{LAA1F}) and (\ref{LAB1F}). If we take this to be the vacancy diffusion coefficient in the perfect solvent lattice, i.e. $4a^2_{Al} \omega_0$, we then obtain
\begin{equation}
\sigma _V=\frac{2\omega_2}{\omega_0}\times \frac{[(3+7F)\omega^{\star}_4+7(1-F)(\omega_0-\omega^{\star}_4)]}{\omega_1+\omega_2+7\omega^{\star}_3F}.
\end{equation}
We proceed to show the results obtained by direct application of the previous theory, to the study of the diffusion of impurities in dilute alloys mediated by a vacancy mechanism.

\section{Results}
\label{S6}

We present our numerical results for $NiAl$ and $AlU$ systems. The interatomic interactions are represented by suitable EAM potentials \cite{PAS08,PAS11,PAS12} for binary systems. For $AlU$, the cross potential has been fitted taking into account the available first principles data \cite{PAS08,PAS12}. Lattice parameters, formation energies and bulk modulus for each intermetallic compound are well reproduced. We obtain the equilibrium positions of the atoms by relaxing the structure via the conjugate gradients technique. The lattice parameters that minimize the crystal structure energy are respectively $a_{Ni}=3.52\, $\AA \, and $a_{Al}=4.05\, $\AA \, for $Ni$ and $Al$ solvents. For all the calculations we used a christallyte of 2048 atoms, eventually including one substitutional $Al$ atom in $Ni$ and one substitutional $U$ atom in $Al$ bulk and a single vacancy in both defective systems. The current calculations have been performed at $T=0K$. In this case, the entropic barrier is ignored. Our calculations are carried out at constant volume, and therefor the enthalpic barrier $\Delta H=\Delta U +p\Delta V$ is equal to the internal energy barrier $\Delta U$.
 
In Table \ref{T4}, we present our results for the vacancy formation energy ($E^V_f$) in pure hosts of $Ni$ and $Al$ calculated as,
\begin{equation}
 E^V_f=E(N-1)+E_c-E(N),
\end{equation}
where, $E(N)$ for the perfect lattice of $N$ atoms, $E(N-1)$ is the energy of the defective system, and $E_c$ the cohesion energy. 
The migration barrier of the vacancy in perfect lattice ($E^V_m$), is calculated with the Monomer method \cite{RAM06}, and the activation energy $E_Q$ as,
\begin{equation}
E_Q=E^V_f+E^V_m.
\end{equation}
\vspace{-1.cm}
\begin{table}[ht] 
\begin{center}
\caption{Energies and lattice parameters for the pure $Al$ and $Ni$ f.c.c. lattices. The first column specifies the metal, vacancy formation energy $E^V_f(eV)$ are shown in the second column. The third column displays the migration energies $E^V_m$, calculated from the Monomer method \cite{RAM06}. In the forth column we show the lattice parameter $a_{A}$(\AA). The last column displays the activation energy $E_Q(eV)$.}
\label{T4}
\begin{tabular}{c|ccccc} 
\hline 
\, Reference \, & \, Latt. \, & \, $E^V_f(eV)$ \, & \, $E^V_m (eV)$  \, & \, $a_{A}$(\AA) \, & \, $E_Q(eV)$\\
\hline 
\hline 
\, Present work \, & \, $Al$ \, & \, 0.649 \, & \, 0.65 \, & \, 4.05 \, & \, 1.299 \\
\, \cite{MIS09} \, & \, $Al$ \, & \, 0.675 \, & \, 0.63 \, & \, 4.05 \, & \, 1.305 \\
\hline 
\, present work \, & \, $Ni$ \, & \, 1.56 \, & \, 0.85 \, & \, 3.52 \, & \, 2.41 \\
\, \cite{ZAC12} \, & \, $Ni$ \, & \, 1.40 \, & \, 1.28 \, & \, 3.52 \, & \, 2.65 \\
\hline
\end{tabular}
\end{center}
\end{table} 

For the case of a diluted alloy, we may consider the presence of solute vacancy complexes, $C_n=S+V_n$ in which $n=1^{st}, 2^{nd}, 3^{rd},\dots$ (see the insets in 
Fig. \ref{T5}) indicates that the vacancy is a $n-$nearest neighbors of the solute atom
$S$. 
The binding energy between the solute and the vacancy for the complex $C_n=S+V_n$ in a f.c.c. matrix of $N$ atomic sites is obtained as,
\begin{equation}
E_b=\left\{ E(N-2,C_n) + E(N)\right\} - \left\{ E(N-1,V) + E(N-1,S)\right\} ,
\label{Ebnn}
\end{equation}
where $E(N-1,V)$ and  $E(N-1,S)$  are the energies  of a crystallite containing ($N-1$) atoms of solvent $A$ plus one vacancy $V$, and one solute atom $S$ respectively, while $E(N-2,C_n)$ is the energy of the crystallite containing ($N-2$) atoms of $A$ plus one solute vacancy complex $C_n=S+V_n$. With the sign convention used here $E_b <0$ means attractive solute-vacancy interaction, and $E_b>0$ indicates repulsion. 

\noindent We calculate the migration energies $E_m$ using the Monomer Method \cite{RAM06}, a static technique to search the potential energy surface for saddle configurations, thus providing detailed information on transition events. The Monomer computes the least local curvature of the potential energy surface using only forces. The force component along the corresponding eigenvector is then reversed (pointing ``up hill"), thus defining a pseudo force that drives the system towards saddles. Both, local curvature and configuration displacement stages are performed within independent conjugate gradients loops. The method is akin to the Dimer one from the literature \cite{HEN01}, but roughly employs half the number of force evaluations which is a great advantage in ab-initio calculations. 

\noindent Binding energies in Tables \ref{T5} and \ref{T6} are displayed, respectively for $NiAl$ and $AlU$. Relative to the $NiAl$ system, a weak energy interaction, $E_b$, between that vacancy and solute can be observed in almost all the nearest neighbor configurations. Also, a weak attractive interaction exists between the vacancy an the $Al$ solute atom only at $1^{st}$ and $4^{th}$ nearest neighbor configurations, while is repulsive for the rest of the pairs. The same behavior is observed for the $AlU$ system but in this case, the binding energy of the pair ($U+V$) at first neighbor position, is strongly attractive. Tables \ref{T5} and \ref{T6} also display the different type of solute vacancy complex $C_n=S+V_n$ with its binding energies $E_b$. Also, the same tables, depict the possibles configurations and jumps that involve the corresponding $C_n=S+V_n$ complex with the corresponding jump frequencies. This jumps imply in a migration of the vacancy whose energies are shown for the direct as well as for the reverse jumps. Relative to the migration barriers, we see that, for $NiAl$ and from Tables \ref{T5}, the vacancy migration barriers $E^{\leftarrow}_m$ are close to that in the perfect lattice $E^V_m=0.85eV$. 

\clearpage
\begin{table}[htdp] 
\begin{center}
\caption{ Jumps and frequencies in $NiAl$. The first column denotes $C_n=S+V_n$ where $V_n$ means that the vacancy is $n$ nearest neighbor of the solute. Binding energy $E_b$ is shown in the second column. The jumps are depicted in the third column, while the forth column describes the jump frequency  $\omega_i$ and the configurations involved in each jump. Migration energies $E_m$ for direct and reversed jumps are written in the fifth and sixth column respectively.}
\label{T5}
\begin{tabular}{cccccc} 
\hline 
\, $C_n=S+V_n$ \, & \, $E_b(eV)$ &Config.($F_n$)& \, $\omega_i $ \, & \, $E^{\rightarrow}_m(eV)$ \, & \, $E^{\leftarrow}_m(eV)$ \, \\
\hline 
\, $C_1$ \, & \, -0.06 \, & 
\begin{minipage}{2.8cm} 
 \includegraphics[width=2.8cm]{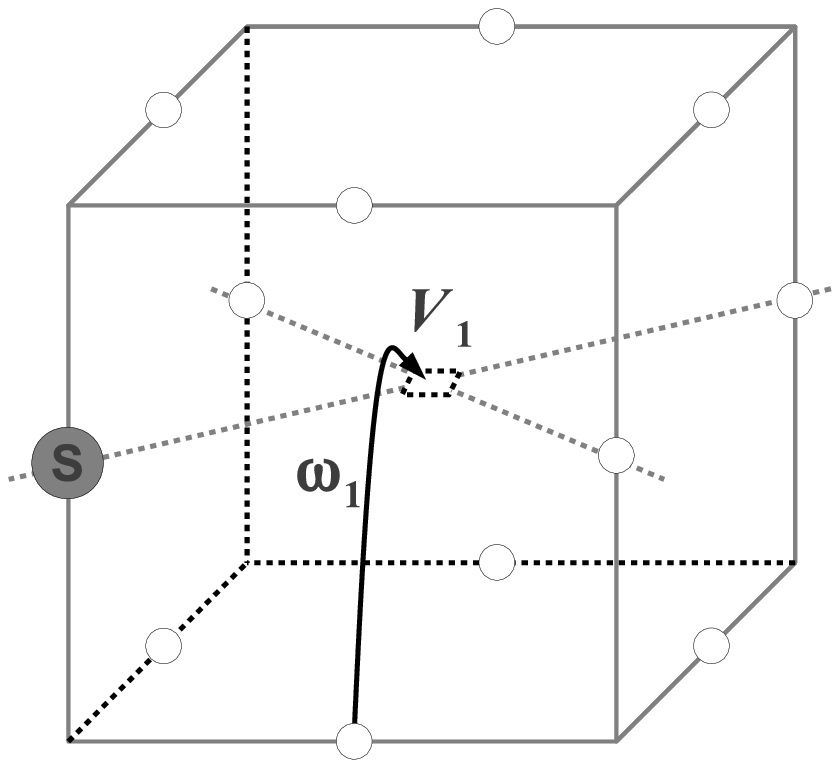} 
\end{minipage}&
\, $\xymatrix{
C_1  \ar@<0.5ex>[r]^{\omega_{1}}
& C_1 \ar@<0.5ex>[l]^{\omega_{1}} }
$ \, & \, 0.94 \, & \, 0.94 \,  \\
\, $C_{1S}$ \, & \, -0.06 \, & 
\begin{minipage}{2.8cm} 
 \includegraphics[width=2.8cm]{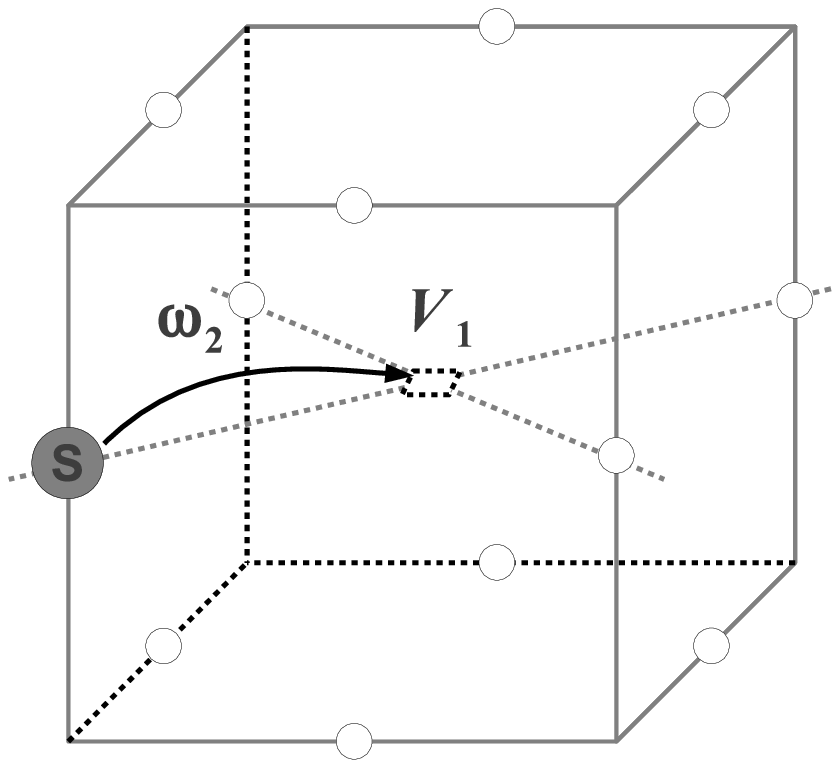} 
\end{minipage}&
\, $\xymatrix{
C_{1S}  \ar@<0.5ex>[r]^{\omega_{2}}
& C_{1S} \ar@<0.5ex>[l]^{\omega_{2}} }
$ \, & \, 0.85 \, & \, 0.85 \, \\
\, $C_2$ \, & \, 0.03 \, & 
\begin{minipage}{2.8cm} 
 \includegraphics[width=2.8cm]{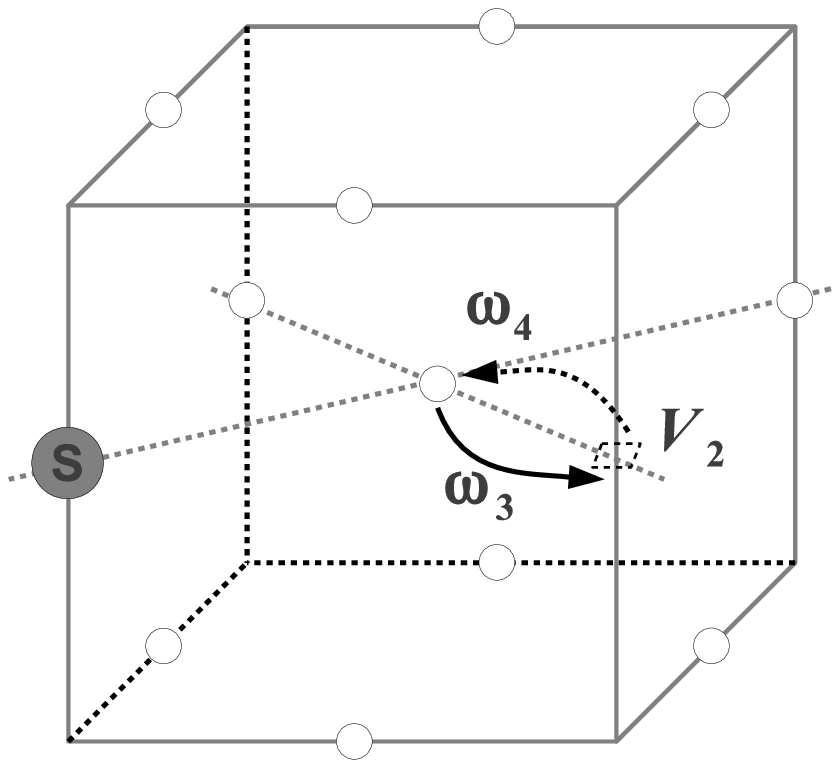} 
\end{minipage}&
\, $\xymatrix{
C_1  \ar@<0.5ex>[r]^{\omega_{3}}
& C_2 \ar@<0.5ex>[l]^{\omega_{4}} }
$ \, & \, 0.85 \, & \, 0.77 \, \\
\, $C_3$ \, & \, 0.03 \, & 
\begin{minipage}{2.8cm} 
\includegraphics[width=2.8cm]{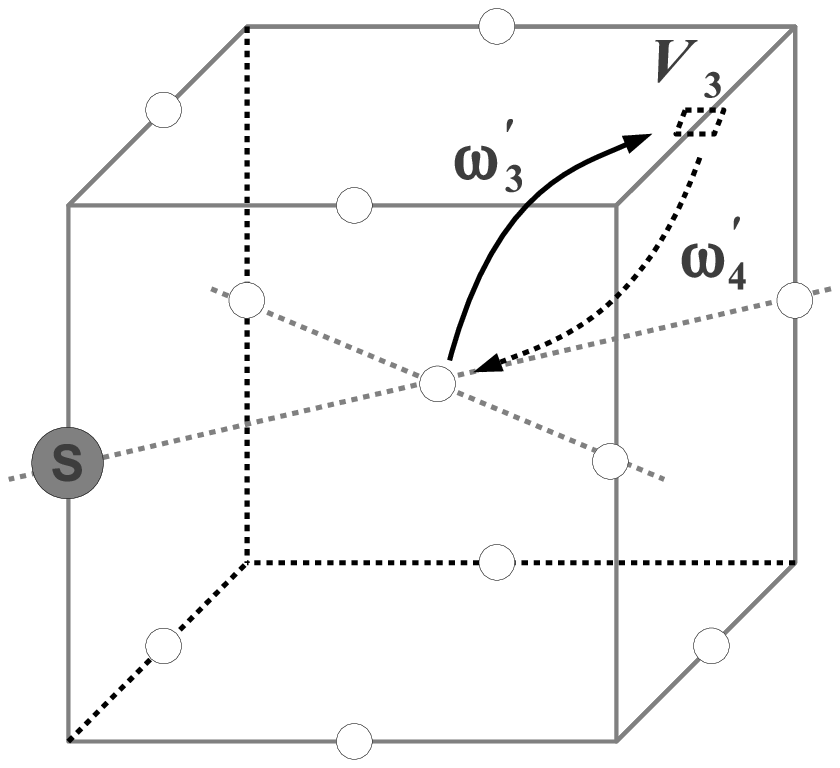} 
\end{minipage}&
\, $\xymatrix{
C_1  \ar@<0.5ex>[r]^{\omega^{\prime}_{3}}
& C_3 \ar@<0.5ex>[l]^{\omega^{\prime}_{4}} }
$ \, & \, 0.86 \, & \, 0.77 \, \\
\, $C_4$ \, & \, -0.001 \, & 
\begin{minipage}{2.8cm} 
\includegraphics[width=2.8cm]{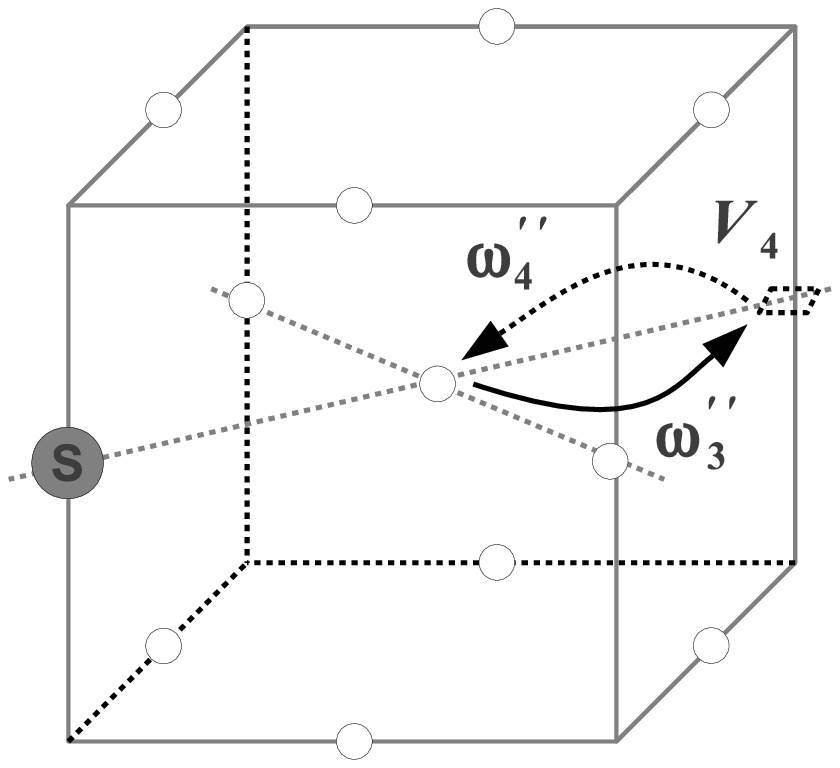} 
\end{minipage}&
\, $\xymatrix{
C_1  \ar@<0.5ex>[r]^{\omega^{\prime\prime}_{3}}
& C_4 \ar@<0.5ex>[l]^{\omega^{\prime\prime}_{4}} }$ \, & \, 0.82 \, & \, 0.76 \, \\
\, $C_5$ \, & \, $-0.001$ \, & 
\begin{minipage}{2.8cm} 
 \includegraphics[width=2.8cm]{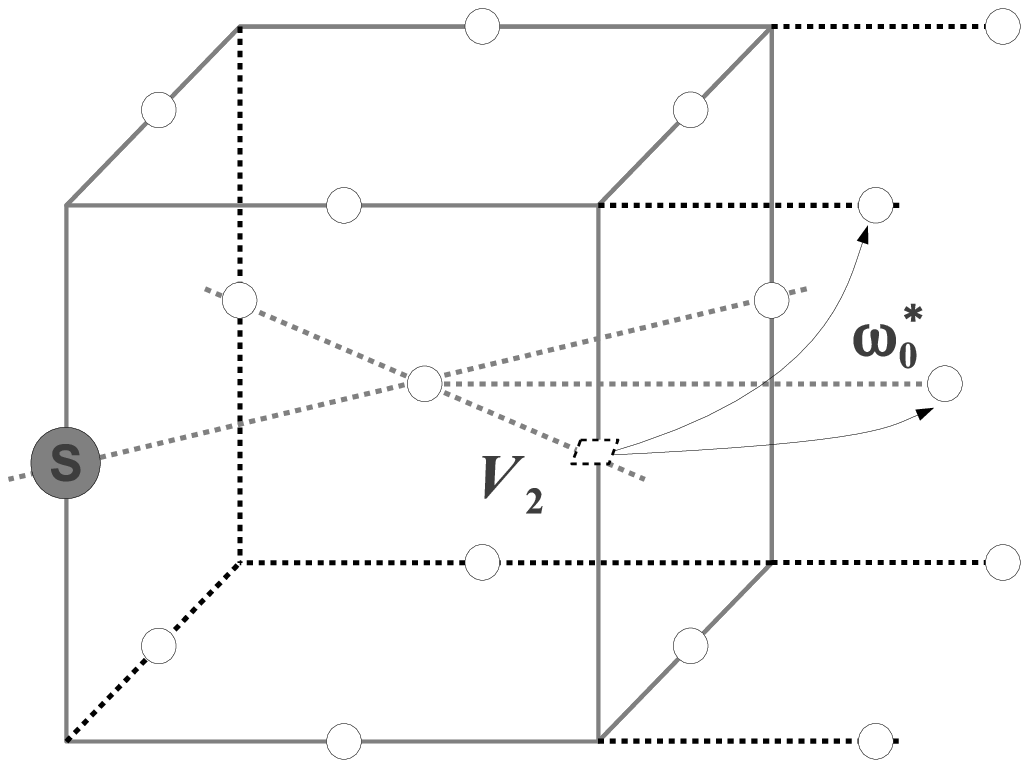} 
\end{minipage}&
\, $\xymatrix{
C_2  \ar@<0.5ex>[r]^{\omega^{\star}_{0}}
& C_5 \ar@<0.5ex>[l]^{\omega^{\star}_{0}} }$ \, & \, 0.84 \, & \, 0.87 \, \\
\, $C_6$ \, & \, $-0.001$ \, & 
\begin{minipage}{2.8cm} 
\includegraphics[width=2.8cm]{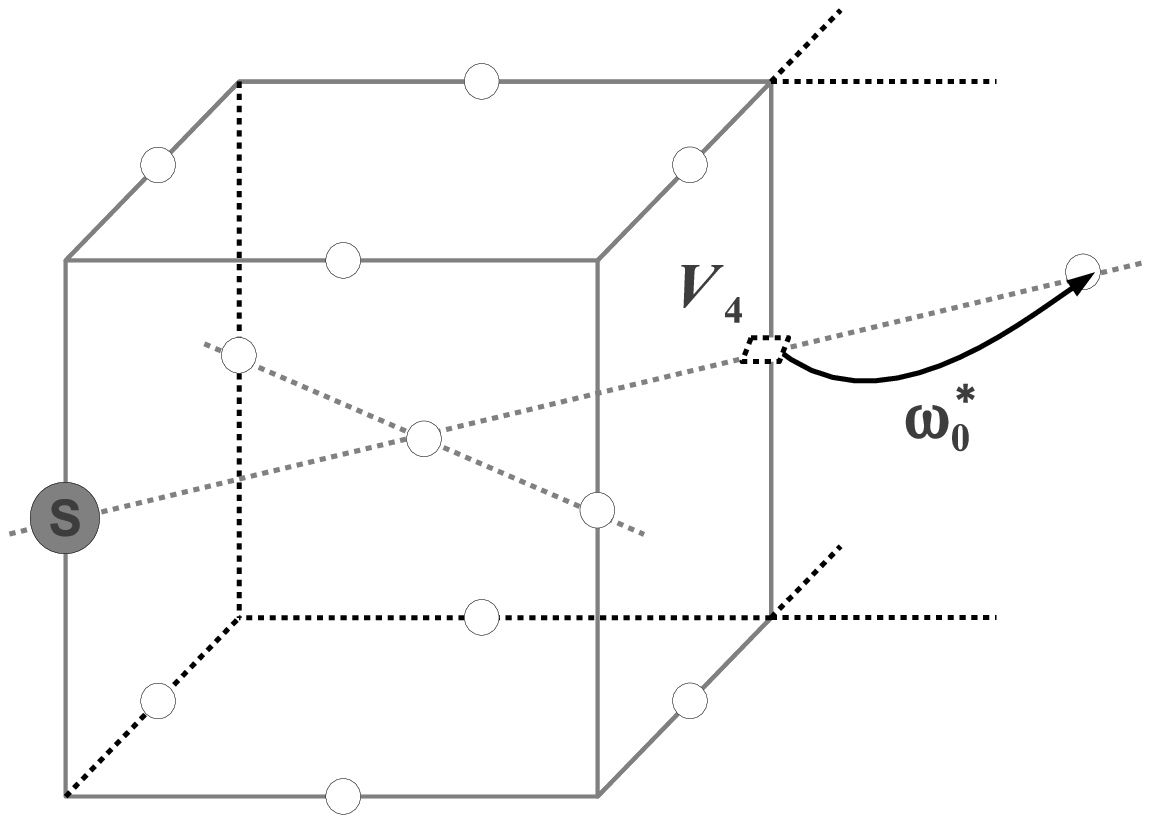} 
\end{minipage}&
\, $\xymatrix{
C_4  \ar@<0.5ex>[r]^{\omega^{\star}_{0}}
& C_6 \ar@<0.5ex>[l]^{\omega^{\star}_{0}} }$ \, & \, 0.84 \, & \, 0.84 \, \\
\, $C_7$ \, & \, $-0.001$ \, & 
\begin{minipage}{2.8cm} 
\includegraphics[width=2.8cm]{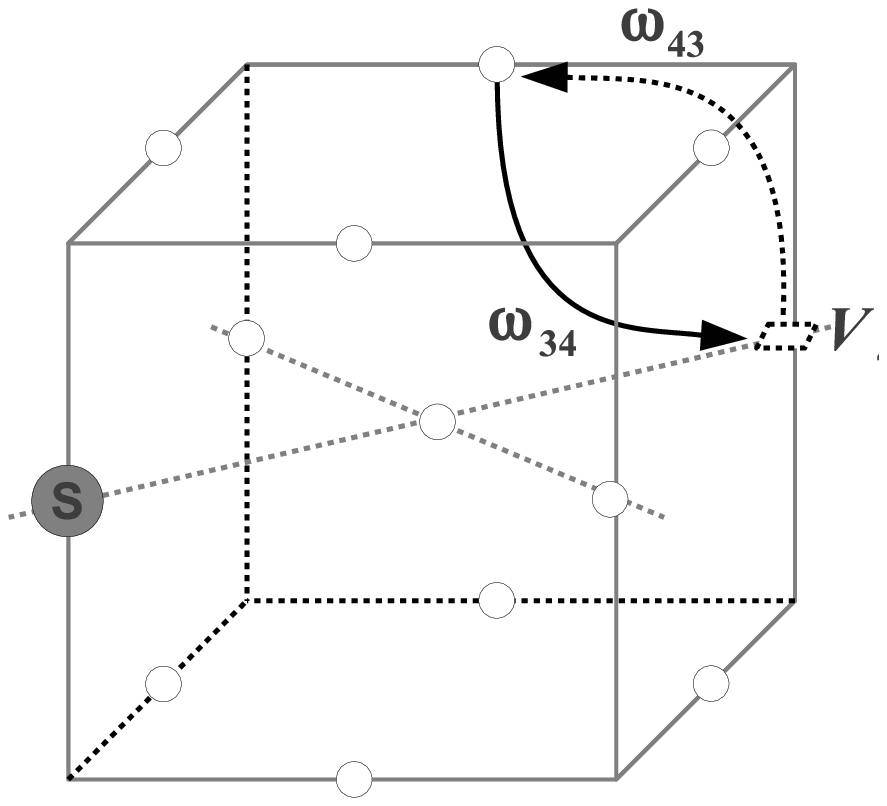} 
\end{minipage}&
\, $\xymatrix{
C_2  \ar@<0.5ex>[r]^{\omega_{43}}
& C_5 \ar@<0.5ex>[l]^{\omega_{34}} }$ \, & \, 0.84 \, & \, 0.87 \, \\
\hline
\end{tabular}
\end{center}
\end{table}

\clearpage

\begin{table}[htdp] 
\begin{center}
\caption{Jumps and frequencies in $AlU$. The columns description is the same as in Table \ref{T5}.}
\label{T6}
\begin{tabular}{cccccc} 
\hline 
\, $C_n=S+V_n$ \, & \, $E_b(eV)$ &Config.($F_n$)& \, $\omega_i $ \, & \, $E^{\rightarrow}_m(eV)$ \, & \, $E^{\leftarrow}_m(eV)$ \, \\
\hline 
\, $C_1$ \, & \, -0.139 \, & 
\begin{minipage}{3.0cm} 
 \includegraphics[width=3.0cm]{FIG5.eps} 
\end{minipage}&
\, $\xymatrix{
C_1  \ar@<0.5ex>[r]^{\omega_{1}}
& C_1 \ar@<0.5ex>[l]^{\omega_{1}} }
$ \, & \, 0.81 \, & \, 0.81 \, \\
\, $C_{1S}$ \, & \, -0.139 \, & 
\begin{minipage}{3.0cm} 
 \includegraphics[width=3.0cm]{FIG6.eps} 
\end{minipage}&
\, $\xymatrix{
C_{1S}  \ar@<0.5ex>[r]^{\omega_{2}}
& C_{1S} \ar@<0.5ex>[l]^{\omega_{2}} }
$ \, & \, 0.48 \, & \, 0.48 \,  \\
\, $C_2$ \, & \, 0.004 \, & 
\begin{minipage}{3.0cm} 
 \includegraphics[width=3.0cm]{FIG7.eps} 
\end{minipage}&
\, $\xymatrix{
C_1  \ar@<0.5ex>[r]^{\omega_{3}}
& C_2 \ar@<0.5ex>[l]^{\omega_{4}} }
$ \, & \, 0.61 \, & \, 0.47 \, \\
\, $C_3$ \, & \, 0.037 \, & 
\begin{minipage}{3.0cm} 
 \includegraphics[width=3.0cm]{FIG8.eps} 
\end{minipage}&
\, $\xymatrix{
C_1  \ar@<0.5ex>[r]^{\omega^{\prime}_{3}}
& C_3 \ar@<0.5ex>[l]^{\omega^{\prime}_{4}} }
$ \, & \, 0.65 \, & \, 0.48 \,  \\
\, $C_4$ \, & \, 0.019 \, & 
\begin{minipage}{3.0cm} 
 \includegraphics[width=3.0cm]{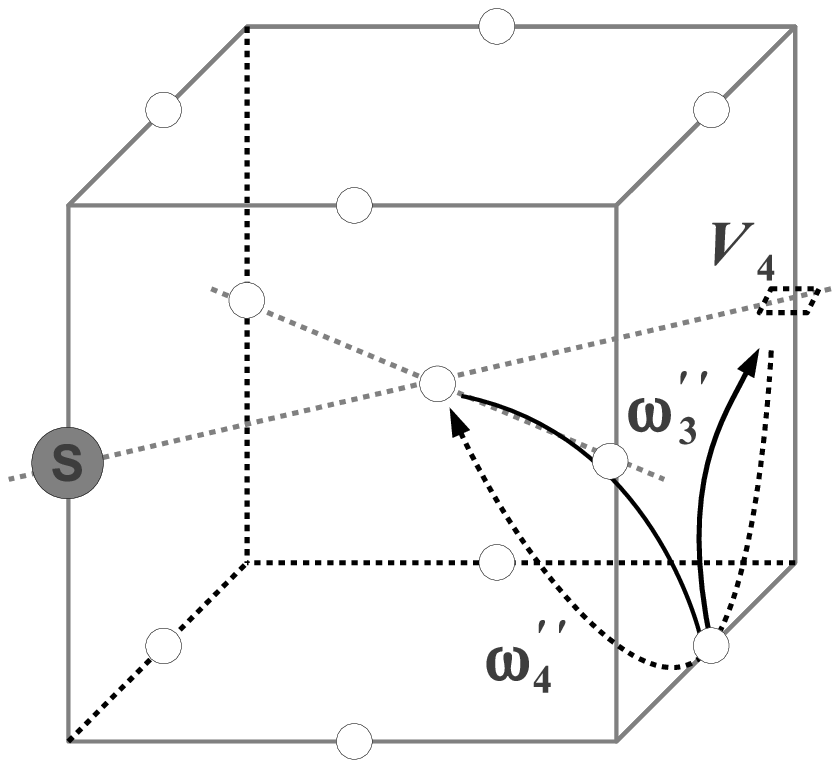} 
\end{minipage}&
\, $\xymatrix{
C_1  \ar@<0.5ex>[r]^{\omega^{\prime\prime}_{3}}
& C_4 \ar@<0.5ex>[l]^{\omega^{\prime\prime}_{4}} }$ \, & \, 0.73 \, & \, 0.58 \, \\
\, $C_5$ \, & \, 0.015 \, & 
\begin{minipage}{3.0cm} 
 \includegraphics[width=3.0cm]{FIG10.eps} 
\end{minipage}&
\, $\xymatrix{
C_2  \ar@<0.5ex>[r]^{\omega^{\star}_{0}}
& C_5 \ar@<0.5ex>[l]^{\omega^{\star}_{0}} }$ \, & \, 0.59 \, & \, 0.58 \, \\
\, $C_6$ \, & \, -0.003 \, & 
\begin{minipage}{3.0cm} 
 \includegraphics[width=3.0cm]{FIG11.eps} 
\end{minipage}&
\, $\xymatrix{
C_4  \ar@<0.5ex>[r]^{\omega^{\star}_{0}}
& C_6 \ar@<0.5ex>[l]^{\omega^{\star}_{0}} }$ \, & \, 0.63 \, & \, 0.65 \,  \\
\hline
\end{tabular}
\end{center}
\end{table}
\clearpage
For the $AlU$ the migration barriers are quite different from that in perfect lattice for association jumps, except for $c_6$. In comparison with the $NiAl$ case, the jump  $\xymatrix{C_1  \ar@<0.5ex>[r]^{\omega^{\prime\prime}_{3}} & C_4 \ar@<0.5ex>[l]^{\omega^{\prime\prime}_{4}} }$, involves more than one atom, i.e. is a multiple jump as indicated in the figure in Table \ref{T6}. In table \ref{T7}, we show the migration barriers for more distant neighbors pairs than the forth, with the purpose to find out from where the jump frequency are similar to that of the perfect crystal $\omega_0$. 
\begin{table}[htdp] 
\begin{center}
\caption{Jumps beyond the second coordinated shell. The columns denotes the same notation as in Table \ref{T5}, binding energies are shown in the second column. The third column denoted the frequency rate, where the supra indexes  $^{(\perp,\mp)}$ on $\omega_0$ implies vacancy jumps perpendicular to, backward or forward $\mp \hat{x}$ respectively. Migration energies for the direct and backward jumps ares shown in column four and five respectively}
\label{T7}
\begin{tabular}{ccccc} 
\hline 
\, $C_n=S+V_n$ \, & \, $E_b(eV)$ \, & \, $\omega_i $ \, & \, $E^{\rightarrow}_m(eV)$ \, & \, $E^{\leftarrow}_m(eV)$ \, \\
\hline 
\hline
\, $C_7$ \, & \, 0.002 \, &  \, $\stackrel{\omega^{\perp}_{0}} {C_7\rightarrow C_{10}}$
\, & \, 0.61 \, & \, 0.64 \,  \\ 
\, $C_{8}$ \, & \, 0.015 \, & \, $\stackrel{\omega^{-}_{0}}{C_{8}\rightarrow C_{11}}$
\, & \, 0.64 \, & \, 0.61 \,  \\
\, $C_9$ \, & \, 0.002 \, & \, $\stackrel{\omega^{+}_{0}}{C_{12} \rightarrow  C_{12}} $ 
\, & \, 0.61 \, & \, 0.64 \, \\
\hline 
\end{tabular}
\end{center}
\end{table}
\vspace{-1.5cm}
\begin{figure}[h]
\begin{center}
\includegraphics[width=7.0cm]{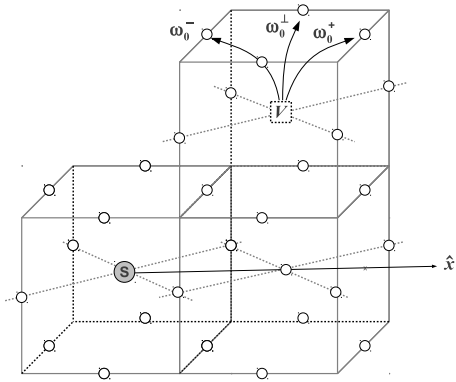}
\vspace{-1.cm}
\caption{Vacancy jumps beyond the second coordinated shell. The supra indexes  $^{(\perp,\mp)}$ on $\omega_0$ implies vacancy jumps perpendicular to, backward or forward $\mp \hat{x}$ respectively.}
\label{FIG14}
\end{center}
\end{figure}
In order to obtain the jump frequencies, we assume that the jumps are thermally activated and then the frequencies $\omega_i$ can be expressed as,
\begin{equation}
\omega_{i}=\nu_{0}\exp(-E^{\rightarrow}_m/k_BT).
\label{nu0T}
\end{equation}
where $E^{\rightarrow}_m$ are reported in Tables \ref{T5} and \ref{T6} for both systems. For the prefactor in (\ref{nu0T}), we use a constant attempt frequency $\nu _{0}=6\times 10^{12}Hz$, taken from Ref. \cite{MIS09} for pure $Al$. We also use, in terms of the Wert model \cite{WER65}, a temperature dependent attempt frequency \cite{TES12} as, 
\begin{equation}
\nu_{0}(T)=\frac{k_BT}{h},
\label{nuT}
\end{equation}
where $h$ is the Planck constant. Also in Tables \ref{T5} and \ref{T6}, the migration barriers and the corresponding rate frequency for each jump are shown. For both $NiAl$ and $AlU$, Table \ref{T8} presents the calculated frequencies at two different temperatures. We adopt the Wert model as in Ref. \cite{TES12,ZAC12}, i.e, a temperature dependent pre-exponential factor from(\ref{nu0T}). 
\begin{table}[htdp]
\begin{center}
\caption{Vacancy jump frequencies rate $\omega_i$ calculated with a temperature dependent attempt frequency $\nu _{0}(T)$, at two  different temperatures in $NiAl$   and $AlU$   alloys. The symbol ($^{\star}$) indicates that we are calculating the effective frequencies $\omega ^{\star}_{3}$ and $\omega ^{\star}_{4}$ .}
\label{T8}
\begin{tabular}{c|cc|cc} 
\hline 
\, & \,\,\,\,\,\,\,\,\,\, $NiAl$  \, && \, $AlU$ \, &
\tabularnewline 
\hline 
\, &\, $T_1=800K$\, & \, $T_2=1700K$\, & \, $T_1=300K$\, & \, $T_2=700K$\, 
\tabularnewline 
\, $\omega_i$ \, & \, $ \omega_i(Hz)$ \, & \, $ \omega_i(Hz)$ \, & \, $ \omega_i(Hz)$\, & \, $ \omega_i(Hz)$\, 
\tabularnewline 
\hline 
\, $\omega_0$ \, & \, $7.36\times 10^{7}$ \, &\, $1.07\times 10^{11}$ \, & \, $1.63\times 10^{2}$ \, &\, $4.25\times 10^{8}$ \, 
\tabularnewline 
\, $\omega_1$ \, &\, $1.99\times 10^{7}$ \, & \, $5.79\times 10^{10}$ \, & \, $1.87\times 10^{-2}$ \, &\, $2.33\times 10^{7}$ \, 
\tabularnewline  
\, $\omega_2$ \, & \, $7.36\times 10^{7}$ \, & \, $1.07\times 10^{11}$ \, & \, $5.17\times 10^{4}$ \, &\, $5.01\times 10^{9}$ \,
\tabularnewline 
\, $\omega^{\star}_3$ \, & \, $7.36\times 10^{7}$ \, & \, $1.06\times 10^{11}$ \, & \, $1.50\times 10^{2}$ \, &\, $3.59\times 10^{8}$ \, 
\tabularnewline 
\, $\omega^{\star}_4$ \, & \, $2.40\times 10^{8}$  \, & \, $1.87\times 10^{11}$ \, & \, $6.24\times 10^{4}$ \, &\, $5.03\times 10^{8}$ \, 
\tabularnewline 
\hline
\end{tabular}
\end{center}
\end{table}

It is clear that the inclusion of $U$ in $Al$ has significant influence on the solvent frequency jumps that the inclusion of $Al$ in $Ni$. This fact may be a consequence of the marked difference between the solute and solvent atomic numbers, $Z_{U}-Z_{Al}=92-13=79$ for $U$ diluted in $Al$, while it is $Z_{Al}-Z_{Ni}=13-28=-15$ for $Al$ in $Ni$. 

Once we have calculated the jump frequencies, then the solute correlation factors $f_S$ and the solvent enhancement factors $b_A$ can be obtained. We present our results in Table \ref{T9}, where we show for different temperatures, the solvent-enhancement factor $b_A^{\star}$ calculated from (\ref{eq:bA}), with $f_0=0.7815$, and the solute-correlation factor $f_S$ from (\ref{eq:FF}), for both $F=1$ and $F\neq 1$ approximations. Also, table \ref{T9}, resumes the jump frequencies ratios calculated according to the five-frequency model of solute-vacancy interaction for pre-exponential factor depending on the temperature.

\begin{table}[h]
\begin{center}
\caption{Solvent enhancement and solute correlated factors for $Ni,Al$ and $Al,U$ at different temperatures, for both $F=1$ and $F\neq 1$ approximations. For the solvent enhancement factor $b_{A}$ (columns two and three), and for the solute correlated factor $f_S$ (columns four and five). The last tree columns describe the jump frequency ratios of the solute$-$vacancy interaction. }
\label{T9}
\begin{tabular}{ccccccccc}
\hline
\hline
\, Alloy \, & \, $T/K$ \, & \, $b^{F=1}_{Ni^*}$ \, &  \, $b^{F\neq1}_{Ni^*}$  \, & \, $f^{F=1}_{Al}$ \, & \, $f^{F\neq1}_{Al}$ \, & \, $\frac{\omega_2}{\omega_1}$ \, & \, $\frac{\omega^{\star}_3}{\omega_1}$ \, & \, $\frac{\omega^{\star}_4}{\omega_0}$ \, \tabularnewline
\hline 
\, $NiAl$   \, & \, 700 \, &  \, 30.22 \, & \, 24.96 \, & \, 0.78\, & \, 0.68 \, & \, 3.78 \, & \, 3.79 \, & \, 3.28 \, \tabularnewline 
 \, & \, 800 \, & \, 31.24 \, & \, 25.11 \, & \, 0.79 \, & \, 0.69 \, &          \, 3.69 \, & \, 3.69 \, & \, 3.26 \, \tabularnewline 
\,  \, & \, 900 \, & \, 25.47 \, & \, 21.05 \, & \, 0.79 \, & \, 0.70 \, &       \, 3.19 \, & \, 3.19 \, & \, 3.86 \, \tabularnewline 
\,  \, & \, 1000 \, & \, 21.38 \, & \, 18.05 \, & \, 0.79 \, & \, 0.71 \, &      \, 2.84 \, & \, 2.83 \, & \, 2.57 \, \tabularnewline 
\,  \, & \, 1100 \, & \, 18.34 \, & \, 15.75 \, & \, 0.79 \, & \, 0.72 \, &       \, 2.58 \, & \, 2.57 \, & \, 2.36 \, \tabularnewline 
\,  \, & \, 1200 \, & \, 16.00 \, & \, 13.93 \,  & \, 0.80 \, & \, 0.72 \, &     \, 2.38 \, & \, 2.37 \, & \, 2.20 \, \tabularnewline 
\,  \, & \, 1300 \, & \, 14.16  \, & \, 12.46 \, & \, 0.80 \, & \, 0.73 \, &     \, 2.23 \, & \, 2.22 \, & \, 2.07 \, \tabularnewline 
\,  \, & \, 1400 \, & \, 12.66 \, & \, 11.25 \, & \, 0.80 \, & \, 0.73 \, &       \, 2.11 \, & \, 2.09 \, & \, 1.96 \, \tabularnewline 
\,  \, & \, 1500 \, & \, 11.43 \, & \, 10.24 \, & \, 0.80 \, & \, 0.73 \, &       \, 2.01 \, & \, 1.99 \, & \, 1.88 \, \tabularnewline 
\,  \, & \, 1600 \, & \, 10.40 \, & \, 9.37 \, & \, 0.80 \, & \, 0.73 \, &       \, 1.92 \, & \, 1.91 \, & \, 1.81 \, \tabularnewline 
\,  \, & \, 1700 \, & \, 9.52 \, & \, 8.63 \, & \, 0.80 \, & \, 0.74 \, &       \, 1.85 \, & \, 1.84 \, & \, 1.74 \, \tabularnewline 
\hline 
\hline
\, Alloy \, & \, $T/K$ \, & \, $b^{F=1}_{Al^*}$ \, &  \, $b^{F\neq1}_{Al^*}$  \, & \, $f^{F=1}_U$ \, & \, $f^{F\neq1}_U$ \, & \, $\frac{\omega_2}{\omega_1}$ \, & \, $\frac{\omega^{\star}_3}{\omega_1}$ \, & \, $\frac{\omega^{\star}_4}{\omega_0}$ \, \tabularnewline
\hline 
\, $AlU$  \, & \,300\, &\, $9.5\times 10^{3}$ \, & \, $4.4\times 10^{3}$ \, & \, $1.0\times 10^{-2}$ \,& \, $2.9\times 10^{-3}$ \, & \, $2.8\times 10^{5}$\, & \, $8.0\times 10^{2}$ \, & \, $3.8\times 10^{2}$ \, \tabularnewline 
\,  \, & \, 350 \, & \, $3.6\times 10^{3}$\, & \, $1.8\times 10^{3}$ \, & \, $2.2\times 10^{-2}$ \, & \, $6.5\times 10^{-3}$ \, & \, $4.6\times 10^{4}$ \, & \, $2.9\times 10^{2}$ \, & \, $1.6\times 10^{2}$ \,  \tabularnewline 
\,  \, & \, 400 \, & \, $1.8\times 10^{3}$ \, & \, $9.4\times 10^{2}$ \, & \, $3.9\times 10^{-2}$ \, & \, $1.2\times 10^{-2}$ \, & \, $1.2\times 10^{4}$ \, & \, $1.4\times 10^{2}$ \, & \, $8.3\times 10^{1}$ \,\tabularnewline 
\,  \, & \, 450 \, & \, $1.0\times 10^{3}$ \,  & \, $5.6\times 10^{2}$ \, & \, $6.0\times 10^{-2}$ \, & \, $1.9\times 10^{-2}$ \, & \, $4.2\times 10^{3}$ \, & \, $7.7\times 10^{1}$ \, & \, $4.9\times 10^{1}$ \, \tabularnewline 
\,  \, & \, 500 \, & \, $6.4\times 10^{2}$ \, & \, $3.6\times 10^{2}$ \, & \, $8.0\times 10^{-2}$ \, & \, $2.9\times 10^{-2}$ \, & \, $1.8\times 10^{3}$ \, & \, $4.9\times 10^{1}$ \, & \, $3.3\times 10^{1}$ \,\tabularnewline 
\,  \, & \, 550 \, & \, $4.4\times 10^{2}$ \, & \, $2.6\times 10^{2}$ \, & \, 0.11 \, & \, $4.2\times 10^{-2}$ \, & \, $9.3\times 10^{2}$ \, & \, $3.4\times 10^{1}$ \, & \, $2.4\times 10^{1}$ \, \tabularnewline 
\,  \, & \, 600 \, & \, $3.3\times 10^{2}$ \, &  \, $1.9\times 10^{2}$ \, & \, 0.14 \, & \, $6.7\times 10^{-2}$ \, & \, $5.3\times 10^{2}$ \, & \, $2.5\times 10^{1}$ \, & \, $1.8\times 10^{1}$ \, \tabularnewline 
\,  \, & \, 650 \, & \, $2.5\times 10^{2}$ \, &  \, $1.5\times 10^{2}$ \, & \, 0.17 \, & \, $7.8\times 10^{-2}$ \, & \, $3.3\times 10^{2}$ \, & \, $1.9\times 10^{1}$ \, & \, $1.4\times 10^{1}$ \, \tabularnewline 
\,  \, & \, 700 \, & \, $1.9\times 10^{2}$ \, &  \, $1.2\times 10^{2}$ \, & \, 0.20 \, & \, $9.2\times 10^{-2}$ \, & \, $2.1\times 10^{2}$ \, & \, $1.5\times 10^{1}$ \, & \, $1.2\times 10^{1}$ \,  \tabularnewline 
\,  \, & \, 750 \, & \, $1.6\times 10^{2}$ \, &  \, $1.0\times 10^{2}$ \, & \, 0.23 \, & \, 0.11 \, & \, $1.50\times 10^{2}$ \, & \, $1.3\times 10^{1}$ \, & \, $1.0\times 10^{1}$ \,  \tabularnewline 
\hline 
\end{tabular}
\end{center}
\end{table}

\begin{figure}[h]
\begin{center}
\includegraphics[angle=-90,width=8.0cm]{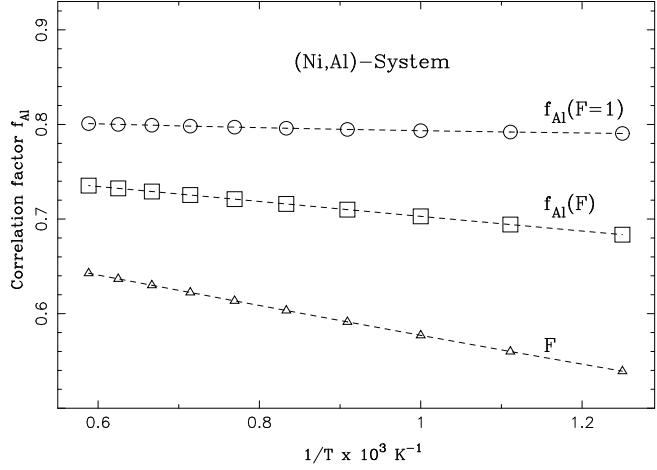}
\vspace{1.5cm}
\caption{Solvent correlation factor  $f_{Al^{\star}}$ in the $NiAl$ system as a function of the temperature for both, $F=1$ (circles) and $F\neq 1$ (squares) approximations. The $F$ factor is denoted with up triangles.}
\label{FIG15}
\end{center}
\end{figure}
\begin{figure}[h]
\begin{center}
\includegraphics[angle=-90,width=8.0cm]{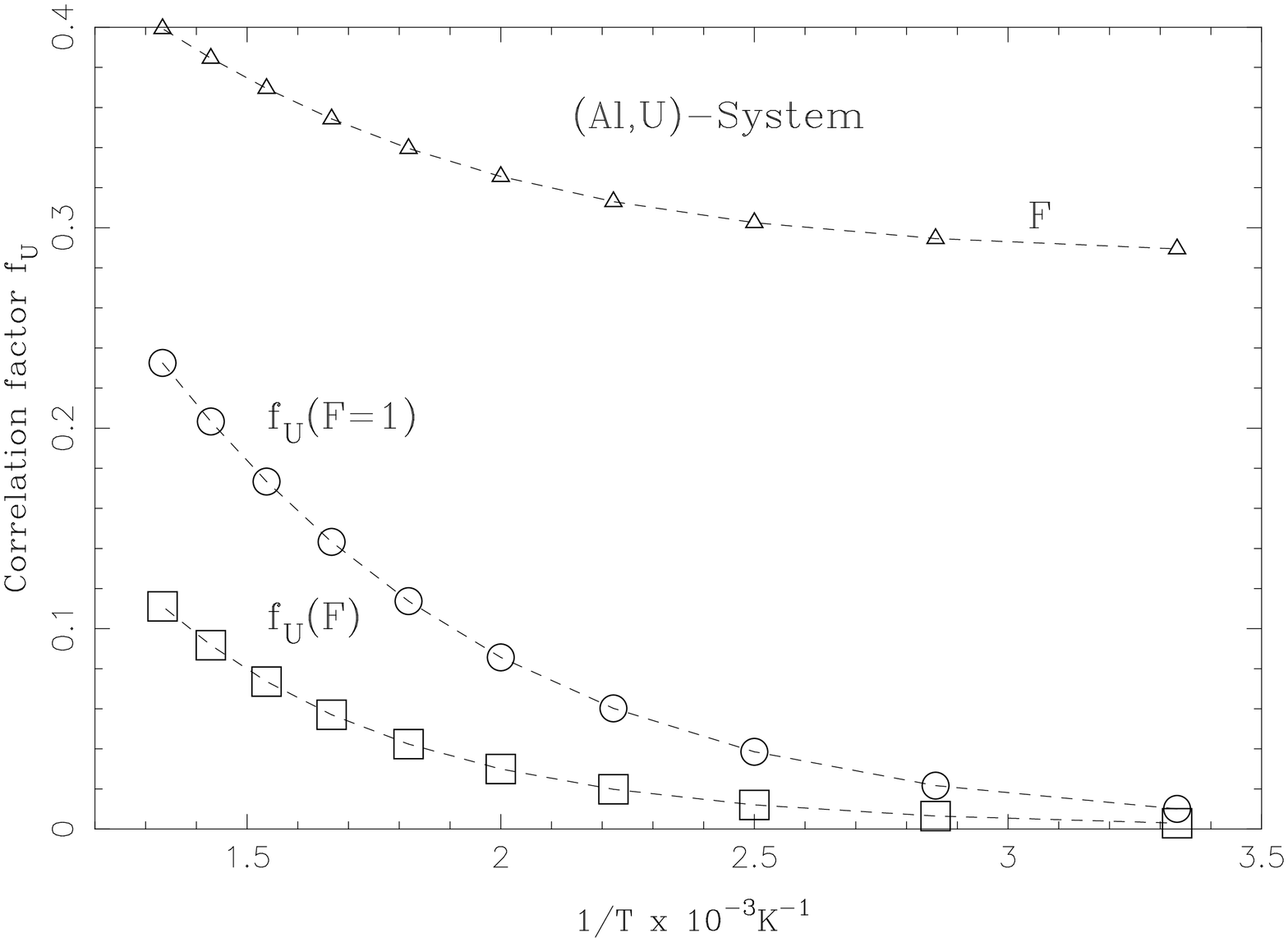}
\vspace{1.5cm}
\caption{Same as figure \ref{FIG15} for the $AlU$ system.}
\label{FIG16}
\end{center}
\end{figure}

The solute-correlation factor ($f_S$) with $T$ and calculated from (\ref{eq:FF1}) and (\ref{eq:FF}) in the $F=1$ and $F\neq 1$ approximations. They are shown in Table \ref{T9} and Figures \ref{FIG15} and \ref{FIG16}, for $NiAl$ and $AlU$ respectively. The factor $F$ obtained from equation (\ref{FW}) is also shown. 

\begin{figure}[h]
\begin{center}
\includegraphics[angle=-90,width=8.0cm]{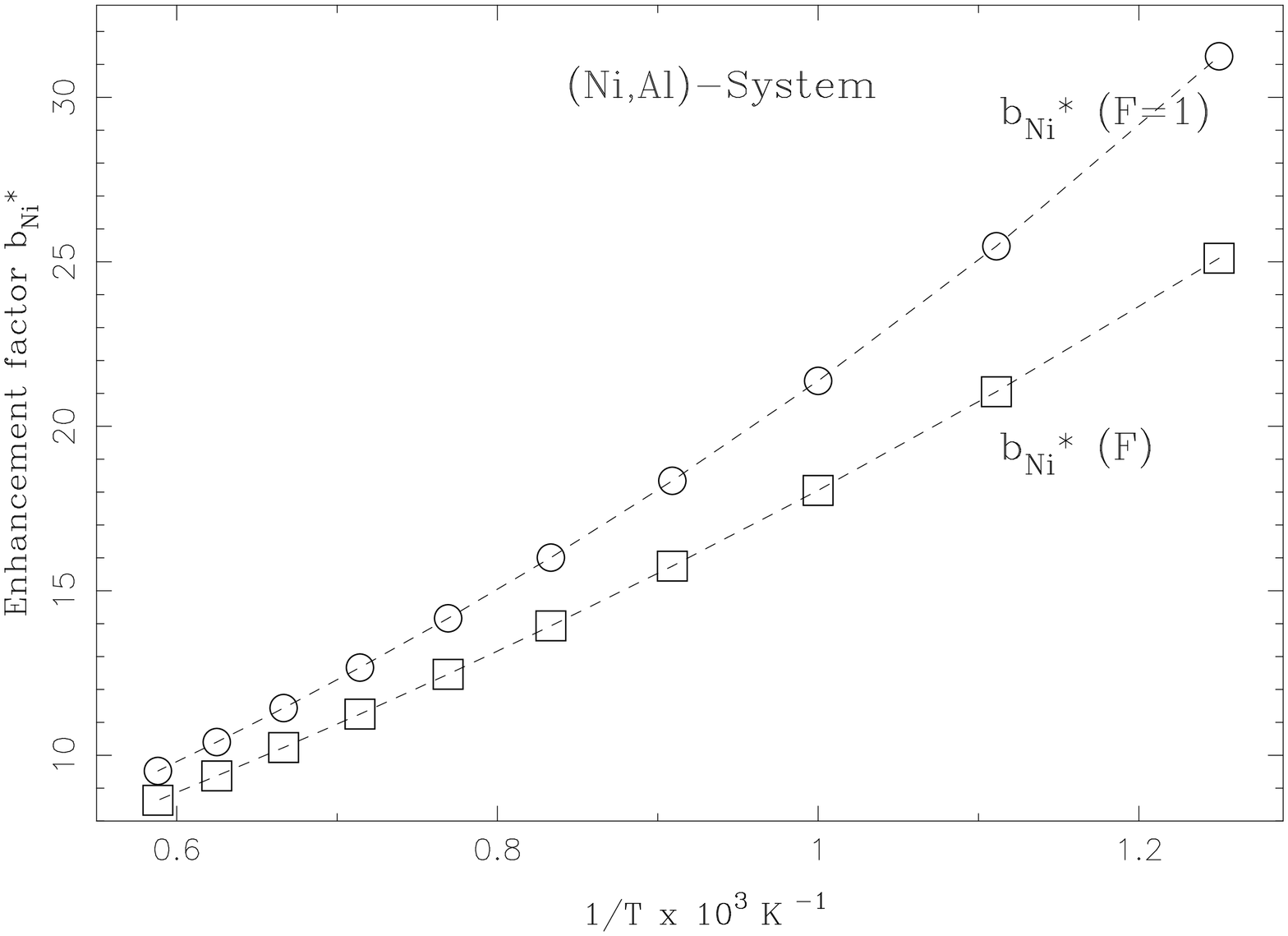}
\vspace{1.5cm}
\caption{Solvent-enhancement factor $b_{Ni}$ in the $NiAl$ system as a function of the temperature for both, $F=1$ (circles) and $F\neq 1$ (squares) approximations.}
\label{FIG17}
\end{center}
\end{figure}

\begin{figure}[h]
\begin{center}
\includegraphics[angle=-90,width=8.0cm]{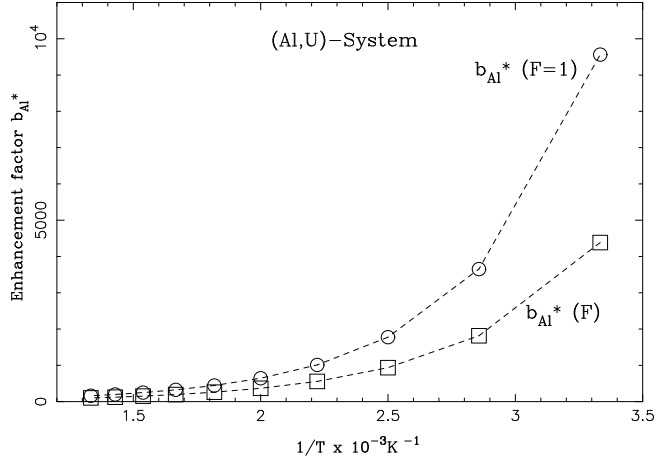}
\vspace{1.5cm}
\caption{Solvent-enhancement factor $b_{Al}$ in the $AlU$ system as a function of the temperature for both, $F=1$ (circles) and $F\neq 1$ (squares) approximations.}
\label{FIG18}
\end{center}
\end{figure}

Concerning to the solvent-enhancement factors, ($b_A$), calculated from (\ref{eq:bA}), the results are shown together with $f_S$ in Table \ref{T9}, also in Figures \ref{FIG17}, \ref{FIG18}, respectively for $NiAl$ and $AlU$, as a function of the temperature. In the Le Claire approximations and for $F\neq 1$, $b_{Ni}$ and $b_{Al}$ are positive with $T$. For $F\neq 1$, $b_A^{\star}$ decrease for both $Ni$ and $Al$ solvents with respect to the Le Claire approximation (i.e., $F=1$). It must be taking into account that this difference will be highly diminished in the diffusion coefficient because the enhancement factor is multiplied by the solute concentration $c_S$, which is low for diluted alloys. 

\begin{figure}[h]
\begin{center}
\includegraphics[angle=-90,width=8.0cm]{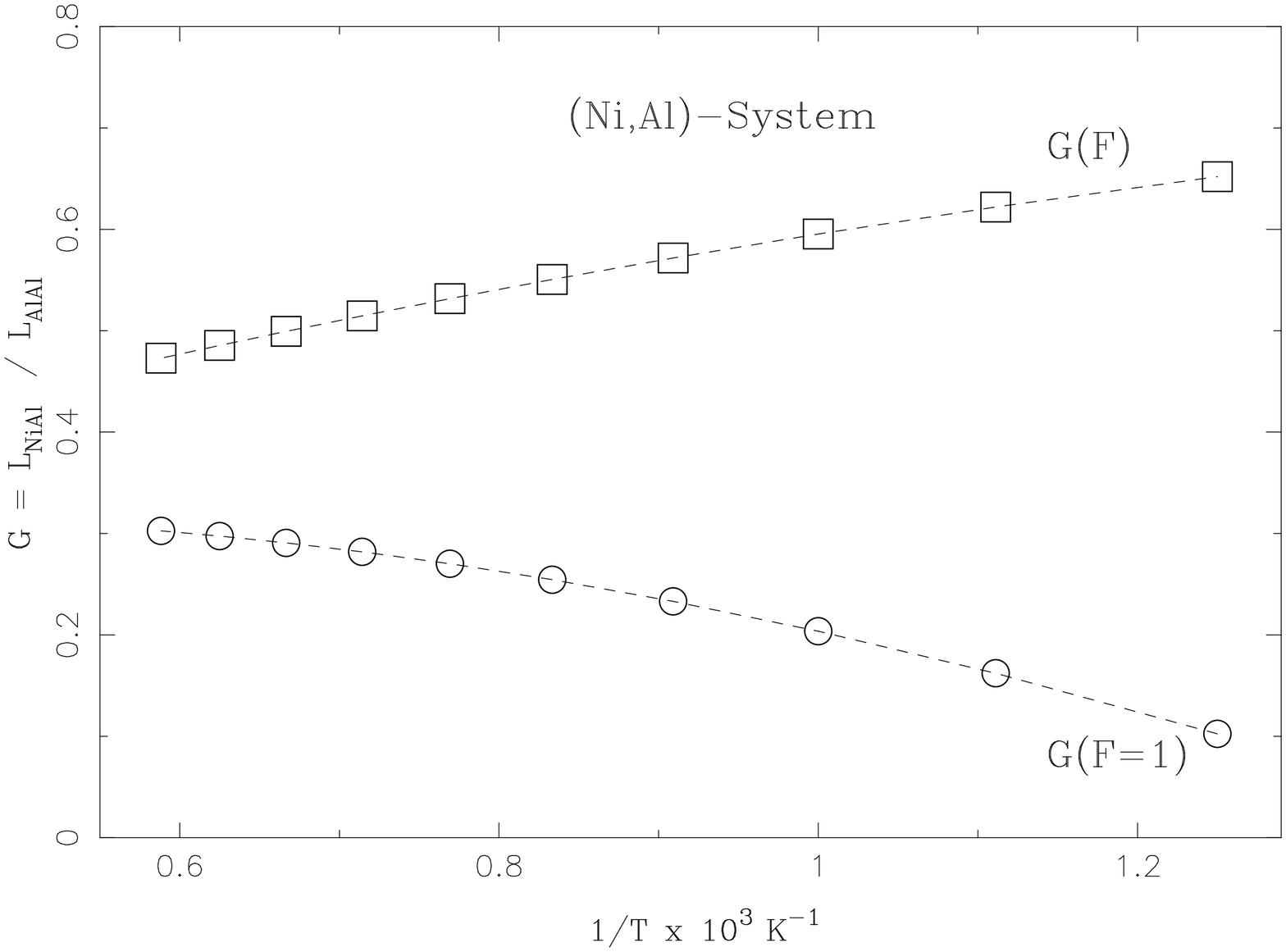}
\vspace{1.5cm}
\caption{Ratio of the vacancy-Onsager coefficients of $Al$ in $Ni$ calculated from eqs.(\ref{LAB1F},\ref{LBB1F}) \textit{vs} $1/T$ for both, $F=1$ (circles) and $F\neq 1$ (squares).}
\label{FIG19}
\end{center}
\end{figure}

\begin{figure}[h]
\begin{center}
\includegraphics[angle=-90,width=8.0cm]{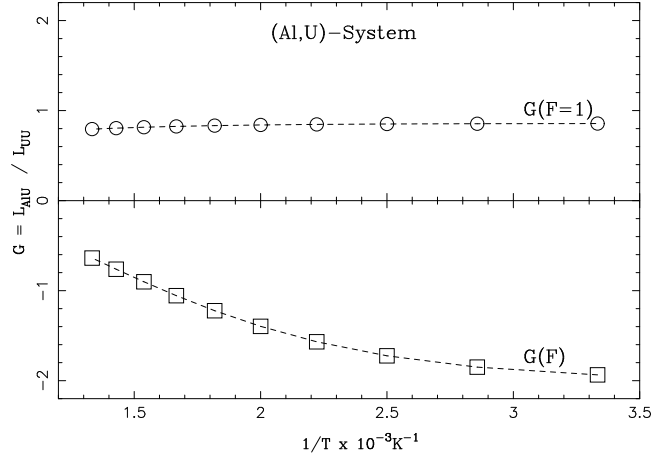}
\vspace{1.5cm}
\caption{Ratio of the vacancy-Onsager coefficients of $U$ in $Al$ calculated from eqs.(\ref{LAB1F},\ref{LBB1F}) \textit{vs} $1/T$ for both, $F=1$ (circles) and $F\neq 1$ (squares).}
\label{FIG20}	
\end{center}
\end{figure}

The Onsager and Diffusion coefficients were calculated assuming a solute mole fraction of $c_S=4.9\times 10^{-4}$, for both alloys, which corresponds to $n_{Al}=4.53\times 10^{19}cm^{-3} \, atoms/cm^3$ for $NiAl$ and $n_U=3.01\times 10^{19}cm^{-3} \, atoms/cm^3$ for $AlU$ system. Once calculated $L_{AS}$ and $L_{SS}$, and following the reasoning in Ref. \cite{CHO11}, we also calculate the vacancy wind coefficient $G=L_{AS}/L_{SS}=-(1+L_{VS}/L_{SS})$. The results are presented in Figures \ref{FIG19} and \ref{FIG20}, for $NiAl$ and $AlU$ systems respectively. We see that if $G<-1$, $L_{VB}$ is positive, then the vacancy and the solute diffuse in the same direction as a complex specie \cite{CHO11}. This transport phenomena could occur in the $AlU$ case, due to the strong binding of the $U+V_1$ pair, while is unlikely to occur for $Al$ in $Ni$ by the opposite argument. The vacancy wind parameter verifies $G>-1$ for $NiAl$ in both, $F=1$ and $F\neq 1$ approximations, while for $Al,U$ the behavior changes drastically depending on the case. If $F=1$, $G$ remains positive, but for $F\neq 1$, $G>-1$ as shown in Fig. \ref{FIG20} in the temperature range $[300-650]^{\circ}C$, this being an indication that a vacancy drag mechanism can occurs for $AlU$.

\begin{figure}[h]
\begin{center}
\includegraphics[angle=-90,width=8.0cm]{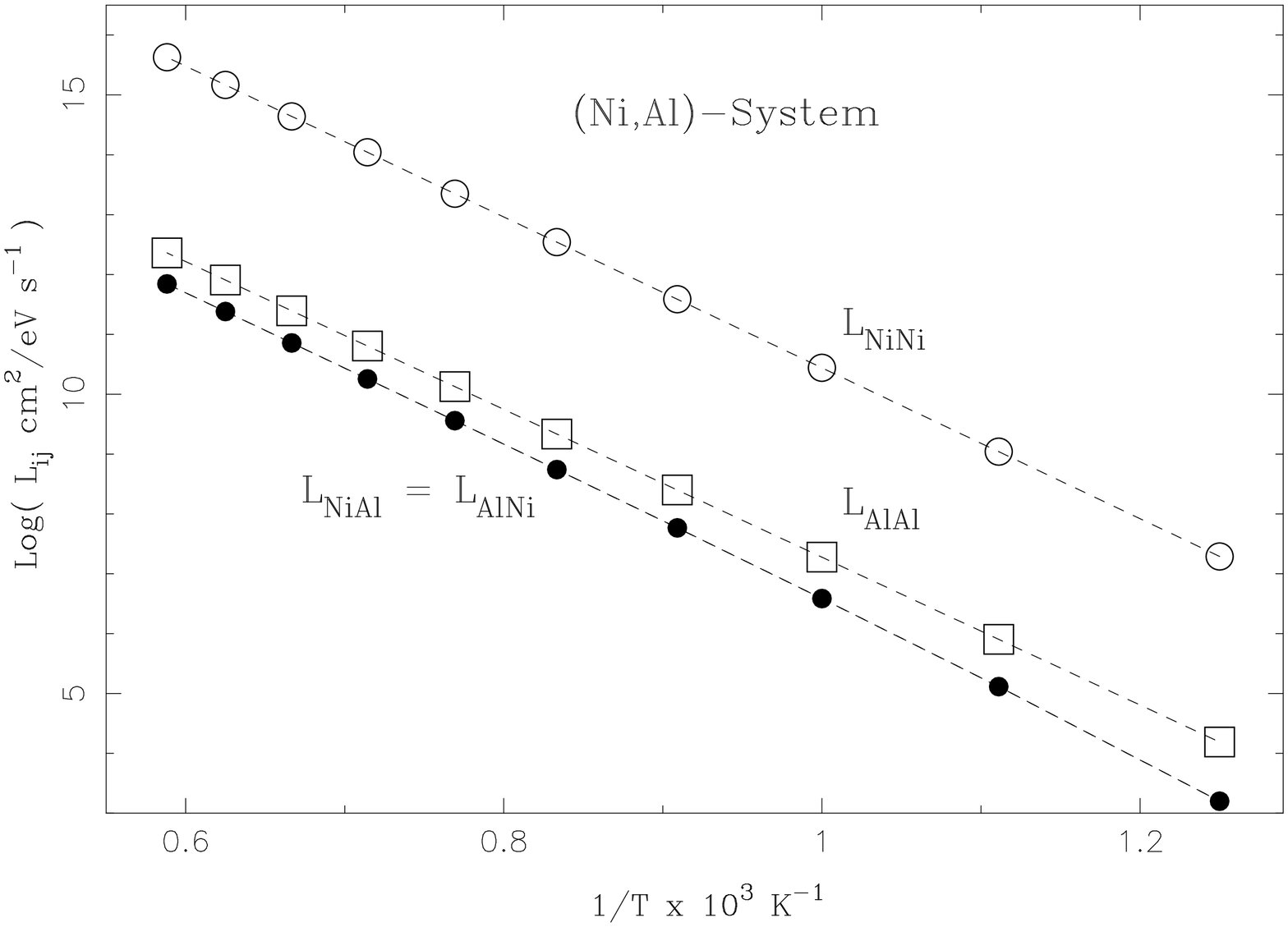}
\vspace{1.5cm}
\caption{Vacancy-Onsager coefficients \textit{vs} $1/T$ for the $NiAl$ system for $F\neq 1$. Squares denote $L_{AlAl}$, empty circles denote $L_{NiNi}$ while $L_{NiAl}$ is described with filled circles.}
\label{FIG21}
\end{center}
\end{figure}

\begin{figure}[h]
\begin{center}
\includegraphics[angle=-90,width=8.0cm]{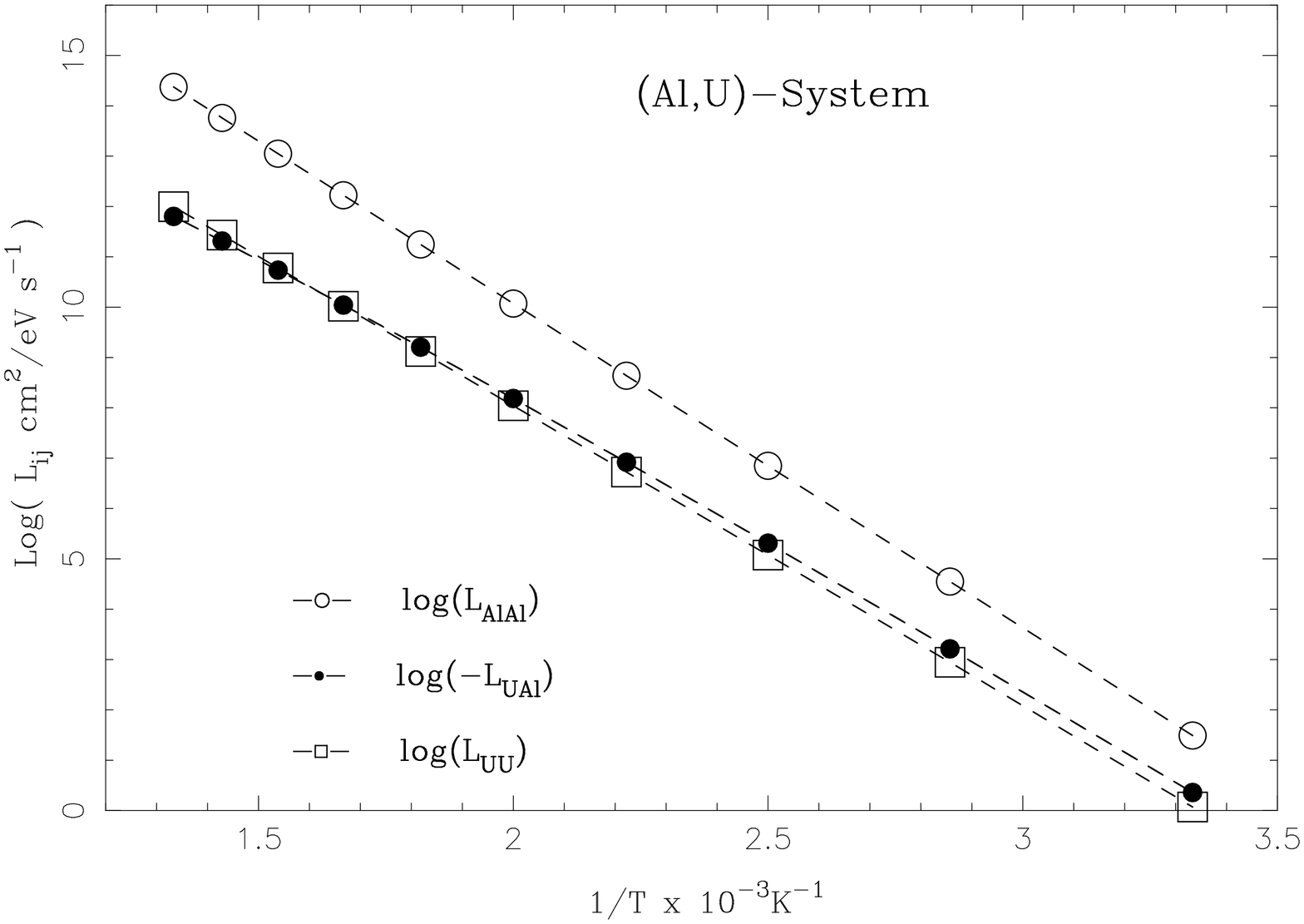}
\vspace{1.5cm}
\caption{Vacancy-Onsager coefficients \textit{vs} $1/T$ for the $AlU$ system  for $F\neq 1$. Squares denote $L_{UU}$, empty circles denote $L_{AlAl}$ while $L_{UAl}$ is described with filled circles.}
\label{FIG22}
\end{center}
\end{figure}

\begin{figure}[h]
\begin{center}
\includegraphics[angle=-90,width=8.0cm]{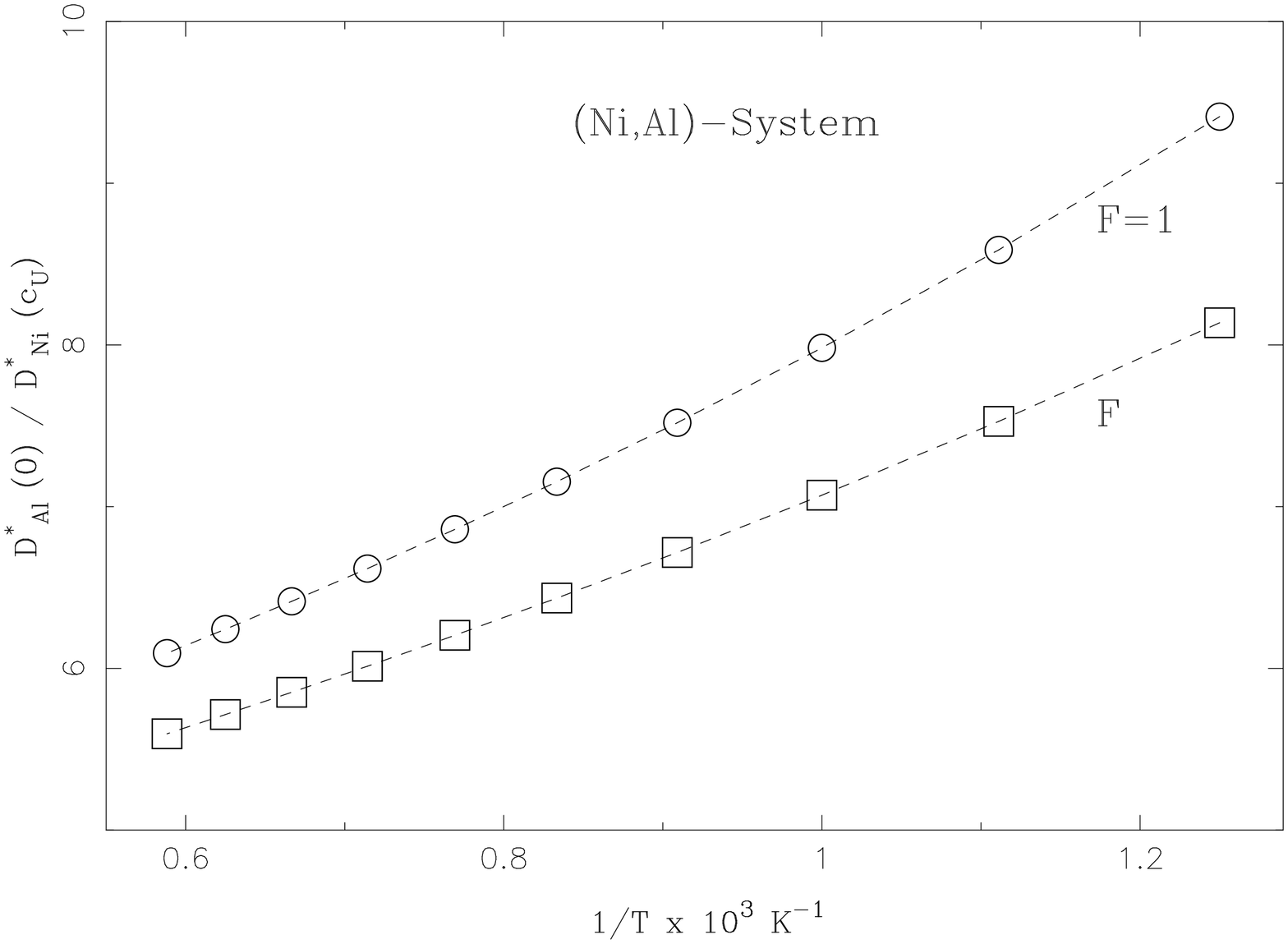}
\vspace{1.5cm}
\caption{Ratio of the tracer diffusion coefficient $D^{\star}_{S}/D^{\star}_{A}$ in $NiAl$ \textit{vs} $1/T$ for both, $F=1$ (circles) and $F\neq 1$ (squares) approximations.}
\label{FIG23}
\end{center}
\end{figure}

\begin{figure}[h]
\begin{center}
\includegraphics[angle=-90,width=8.0cm]{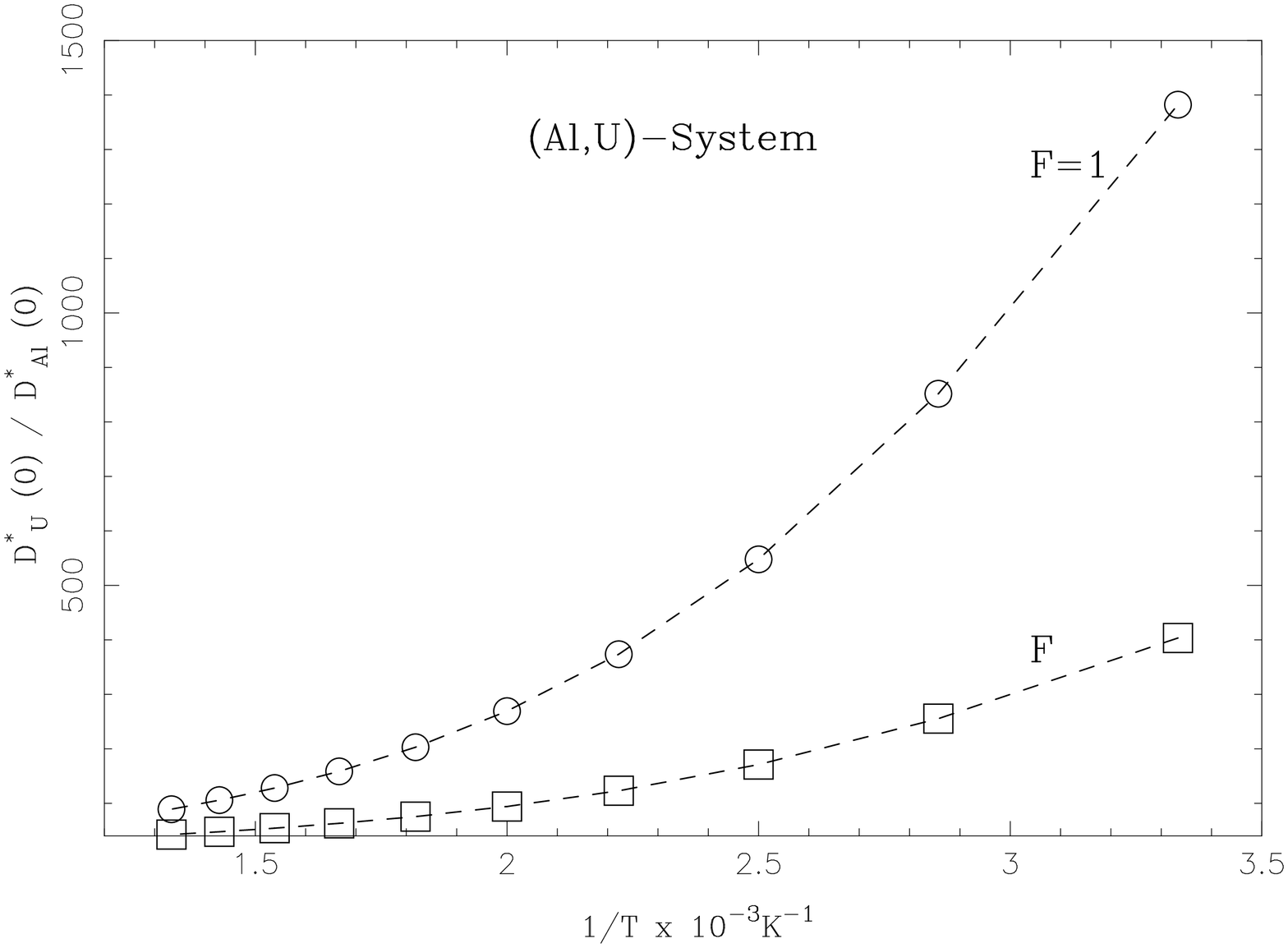}
\vspace{1.5cm}
\caption{Ratio of the tracer diffusion coefficient $D^{\star}_{S}/D^{\star}_{A}$ in $AlU$) \textit{vs} $1/T$ for both, $F=1$ (circles) and $F\neq 1$ (squares) approximations.}
\label{FIG24}
\end{center}
\end{figure}

The full set of $L$-coefficients for $F\neq 1$, are displayed in Figs. \ref{FIG21} and \ref{FIG22}, against the inverse of the temperature for the $NiAl$ and $AlU$ respectively. We see that for the $NiAl$ case the $L$-coefficients follow an Arrhenius behavior, which implies a linear relation between the logarithm of $L$-coefficients against the inverse of the temperature (see Fig. \ref{FIG21}). While for the $AlU$ case, at high temperatures, we can appreciate a slight deviation from the Arrhenius law (see Fig. \ref{FIG22}).  
In Figure \ref{FIG22}, the cross $L_{AlU}=L_{UAl}$ coefficient is negative for all the range of temperature. Now, we are in position to obtain the diffusion coefficients $D{\star}_A(0)$, $D{\star}_B(0)$ and $D_p$, for the paired specie. First, we present the ratio of calculated tracer diffusion coefficients $D^{\star}_{S}/D^{\star}_{A}$ as a function of the inverse of the temperature for the $NiAl$ and $AlU$   in Figures \ref{FIG23} and \ref{FIG24}, respectively. 

\begin{figure}[h]
\begin{center}
\includegraphics[angle=-90,width=8.0cm]{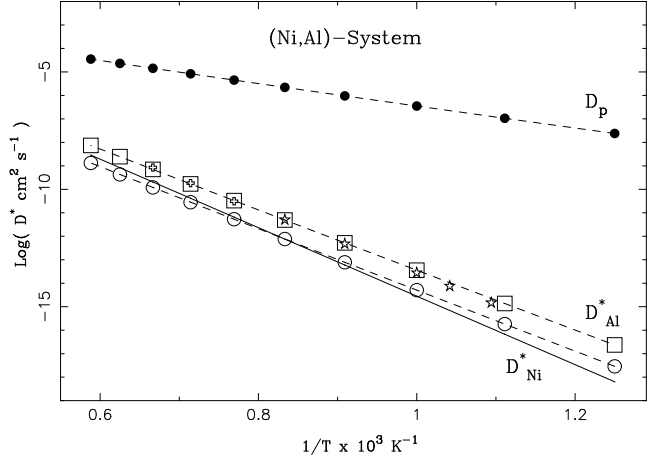}
\vspace{1.5cm}
\caption{Tracer diffusion coefficients of $Al$ ($D^{\star}_{Al}$ in open squares) and $Ni$ ($D^{\star}_{Ni}$ in open circles) in the alloy. Solid line represents the best estimative of the pure $Ni$ self-diffusion coefficient $D^{\star}_L (Ni)$ taken from Campbell work \cite{CAM11}. Available experimental data, for the $Al$ diffusion coefficient in the alloy, are displayed with stars \cite{GUS81} and cruxes \cite{SWA56}.}
\label{FIG25}
\end{center}
\end{figure}

\begin{figure}[h]
\begin{center}
\includegraphics[angle=-90,width=8.cm]{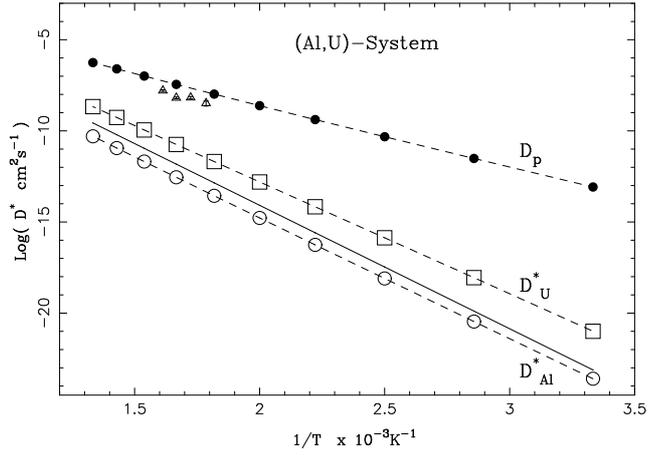}
\vspace{1.5cm}
\caption{Tracer diffusion coefficients of $U$ ($D^{\star}_U$ in open squares) and $Al$ ($D^{\star}_{Al}$ in open circles) in the alloy. Solid line represents the best estimative of the pure $Al$ self-diffusion coefficient $D^{\star}_L (Al)$, taken from Campbell work \cite{CAM11}. Available experimental data, for the $U$ diffusion coefficient in the alloy, are displayed with triangles \cite{HOU71}.}
\label{FIG26}
\end{center}
\end{figure}

The calculated $D^{\star}_A$ and $D^{\star}_S$ for $F\neq 1$, using the equations (\ref{DACB}) and (\ref{DBB2}), are shown in Figures \ref{FIG25} and \ref{FIG26} respectively for $NiAl$ and $AlU$. The diffusion coefficient of the paired specie, $D_p$, calculated from (\ref{Dp}) is also shown. It is import to perform a comparison between theoretical results obtained in present work with reliable experimental data. We have verified that the tracer self diffusion coefficient $D^{\star}_A(c_S)$ for a diluted alloy is practically equal to that for the pure solvent $D^{\star}_A(0)$ (i.e., $D^{\star}_A(c_S) \simeq D^{\star}_A(0)$). Hence, we can test our results for $D^{\star}_A(c_S)$ with the best estimative of the diffusion parameter for pure solvent, $D^{\star}_L (A)$, taken from Campbell et al. \cite{CAM11}. The authors, have been used weighted means statistics to determine consensus estimators which represents best the available experimental data. They use a Gaussian distribution to represent the experimental error used to determine the best estimates of the parameters common to all of the included studies in the parameter $D^{\star}_L (A)$, the self-diffusivity of species $A$ in pure $A$ given in $cm^2s^{-1}$. The best estimate is given through an expression of the form,
\begin{equation}
D^{\star}_L (A)=D^0_A\exp(-Q_A/RT),
\label{DLCONS}
\end{equation}
where $D^0_A$ and $Q_A$ from Ref. \cite{CAM11} are dysplayed in Table \ref{T10} for pure $Ni$ and $Al$, $R=8.314472$ J/mol K is the ideal gas constant and $T$ is the absolute temperature and represented by solid lines in Figures \ref{FIG25} and \ref{FIG26}.
\begin{table}[htdp] 
\begin{center}
\caption{Parameters involved in the expression for the self-diffusion  consensus fit $D^{\star}_L(A)$, where the parameter $A$ indicates $Ni$ or $Al$ hosts. The first column denotes the reference where the values were taken from. The solvent lattice is indicated in the second column. The third and fourth columns denote the preexponential factor $D^0_A$ and the activation energy $Q_A$ for equation (\ref{DLCONS}) respectively. The range of temperatures of the description is referred in column five, while the relative error of the self diffusion coefficient is shown in column six. The last column stands for experimental or theoretical results. The values were taken from Campbell work \cite{CAM11} and references therein.}
\label{T10}
\begin{tabular}{ccccccc} 
\hline 
\, Ref. \, & \, Lattice \, & \, $D^0_A(cm^2s^{-1})$ \, & \, $Q_A(KJ/mol)$ \, & \, $T(^{\circ}C)$ \, & \, error \, & \, type \\
\hline 
\hline
\, \cite{WAZ65a,WAZ65b} \, & \, $Ni$ \, & \, $1.9$ \, &  \, $279.5$ \, & \, $[773-1023]$ \, & \, $16\%$ \, & \, exp. \\ 
\, \cite{MES74} \, & \, $Al$ \, & \, $0.137$ \, &  \, $123.5$ \, & \, $[90-930]$ \, & \, $15\%$ \, & \, exp. \\ 
\, \cite{CAM11} \, & \, $Ni$ \, & \, $1.1$ \, &  \, $279.35$ \, & \, $[769-1667]$ \, & \, $-$ \, & \, $D^{\star}_L (Ni)$ \\ 
\, \cite{CAM11} \, & \, $Al$ \, & \, $0.292$ \, &  \, $129.7$ \, & \, $[357-833]$ \, & \, $-$ \, & \, $D^{\star}_L (Al)$ \\ 
\hline 
\end{tabular}
\end{center}
\end{table}

As can be observed, $D^{\star}_L (A)$ fits perfectly with the values of $D^{\star}_A$ calculated in the present work. For the case of $NiAl$ alloys, in Fig. \ref{FIG25} experimental data of the solute diffusion coefficient are plotted with stars and cruxes respectively for $T=[914-1212]^{\circ}C$ \cite{GUS81} and $T=[1372-1553]^{\circ}C$ \cite{SWA56}. As we can observe the accuracy with the calculated solute diffusion coefficient $D^{\star}_{Al}$ is astonishing, showing that the here employed procedure gives excellent results for calculating the diffusion coefficients in diluted f.c.c. alloys. The diffusion coefficient for the paired specie $Al+V$ in $Ni,Al$ is also shown. 


A little more attention we devote to $AlU$ system. In the literature, we have found experimental values for the $U$ diffusion coefficient at infinite dilution in $Al$ \cite{HOU71}. The authors fit their own experimental results solving numerically the diffusion equation
\begin{equation}
\frac{\partial C(x,t)}{\partial t}=D\frac{\partial ^2C(x,t)}{\partial x^2},
\label{FIT}
\end{equation}
with boundary condition $x=0;\, C(0,t)=S_0$, where $S_0$ is the maximum solubility of the diffusing specie in the alloy ($U \rightarrow Al$). They propose a solution for equation (\ref{FIT}) as,
\begin{equation}
C(x,t)=S_0[1-erf(x/2\sqrt{Dt}].
\end{equation}
In Table \ref{DEXP} we show the results of the fit of experimental data taken from \cite{HOU71}, against we will compare our theoretical results. Also, the authors argue that at infinite dilution the dissolution of precipitates do not disturb the $U$ process diffusion in $Al$. 
\begin{table}[htdp] 
\begin{center}
\caption{ Solubility and diffusion of $U$ in $Al$. $D\times 10^{8}cm^2s^{-1}$($S_0\times 10^{10}$ at).}
\label{DEXP}
\begin{tabular}{ccccc} 
\hline 
\, $T(^{\circ}C)$ \, & \, $1.5\%Wt$ \, & \, $0.75\%Wt$ \, & \, $0.15\%Wt$ \, & \, $500\times 10^{-6} \%Wt$ \,  \\
\hline 
\hline
\, $620$ \, & \, $1.60 \pm 0.20$ \, &  \, $1.5 \pm 0.15$ \, & \, $1.56 \pm 0.15$ \, & \, $1.62 \pm 0.16$ \,  \\ 
\,       \, & \, ($20 \pm 10$) \, &  \, ($25 \pm 15$) \, & \, ($30 \pm 15$) \, & \, ($20 \pm 10$) \,  \\ 
\, $600$ \, & \, $0.78 \pm 0.08$ \, & \, $0.68 \pm 0.07$\, & \, $0.70 \pm 0.15$ \, & \, $0.65 \pm 0.07$ \,  \\
\,       \, & \, ($10 \pm 5$) \, &  \, ($10 \pm 5$) \, & \, ($12 \pm 7$) \, & \, ($15 \pm 5$) \,  \\ 
\, $580$ \, & \, $0.55 \pm 0.12$ \, & \, $0.70 \pm 0.12$ \, & \, $0.44 \pm 0.15$ \, & \, $0.67 \pm 0.10$ \, \\
\,       \, & \, ($15 \pm 7$) \, &  \, ($20 \pm 10$) \, & \, ($6 \pm 4$) \, & \, ($10 \pm 5$) \,  \\ 
\, $560$ \, & \, $0.40 \pm 0.10$ \, & \, $0.35 \pm 0.10$ \, & \, $0.31 \pm 0.10$ \, & \, $0.33 \pm 0.10$ \, \\
\,       \, & \, ($8 \pm 4$) \, &  \, ($10 \pm 5$) \, & \, ($5 \pm 2$) \, & \, ($6 \pm 3$) \,  \\ 
\hline 
\end{tabular}
\end{center}
\end{table}
In Figure \ref{FIG26}, we establish a comparison with experimental data in Table \ref{DEXP} for an Uranium dilution of $500\times 10^{-6} \%Wt$, which corresponds to $C_U=6.57\times 10^{-5}$. We see that, experimental values are in perfect agreement with $D_p$, contrarily to the $NiAl$ for which our calculations and calculations in Ref. \cite{ZAC12} reveal a weak Aluminium-vacany binding, then experimental values of solute diffusion goes with $D^{\star}_{Al}$. 

The diffusion of Uranium into Aluminum was also calculated in a study of the maximum rate of penetration of uranium into aluminum in the temperature range $200-390 ^{\circ}C$ as described in a report from the literature \cite{BIE55}. The maximum values calculated in \cite{BIE55} for the penetration coefficient was, $K_T=x^2/t=0.075,0.50,6.1\times 10^{-6} inch^2/hr$ at temperatures of $200$, $250$ and $390 ^{\circ}C$, respectively. The activation energy $Q$ from the expression $ K=K_0\exp^{-Q/RT}$ in the temperature range $T=[200-390]$ is $Q=14.300$ in calories per mole, $R$ the gas constant in calories per $1/^{\circ}C$ per mole, and $T$ the absolute temperature. $K_O$ is the proportionality constant. The plot $lnK$ vs $1/T$ provides a convenient basis for expressing and comparing penetration coefficients.

Not shown here, but also performed, we recalculate all the microscopical parameters for a crystallite containing $~100$ atoms using classical molecular static technique, including one solute atom and a vacancy at first neighbor sites of the solute. We reproduce all the migration barriers and therefore the jump frequency rates.

In summary, for pure $Ni$ and $Al$ materials, a large amount of experimental data are available in the literature, which have been summarized by Campbell \cite{CAM11} in a best confidence estimation of the self diffusion coefficient. Our calculations for pure hosts match perfectly well with this best estimation when a temperature dependent $\nu _0$ is assumed, although results for a constant value of $\nu _0$, also gives accurate results.

Concerning with diluted alloys, our results are in excellent agreement with experiments \cite{ZAC12,WAZ65a,WAZ65b,MES74} for the tracer diffusion coefficient in $NiAl$. For the diffusion behavior in $AlU$, we only found in the literature the work by Housseau et al. \cite{HOU71}. Our results when compared with the experimental data \cite{HOU71}, suggest that the diffusion behavior is mainly due to a vacancy drag mechanism. 
\section{Concluding remarks}
In summary, We propose a general mechanism based on first principles for obtaining diffusion coefficients.

The flux equations permits to relates the  diffusion coefficients with the Onsager tensor. 
Non  equilibrium thermodynamics allows to write this Onsager coefficients in terms of jump frequencies. In this way we could write expressions for the diffusion coefficients only in terms of microscopic magnitudes, i.e. the jump frequencies. This last ones have been calculated thanks to to the economic static molecular techniques namely the monomer method.

The five frequency model has also been of great utility in order to discriminate the relevant jump frequencies for both the the Le Claire approximation ($F=1$) and one more accurate when $F\ neq 1$ is considered. Hence, we have calculated the full set of phenomenological coefficients from which the full set of diffusion coefficients are obtained through the flux equation.

Although in this work we have performed the treatment for the case of f.c.c. latices where the diffusion is mediated by vacancy mechanism, a similar procedure can be adopted for other crystalline structures or different diffusion mechanism (for example, interstitials).  

We have exemplified our calculations for the particular cases of binary $NiAl$   and $AlU$   f.c.c. diluted alloys. 

When a temperature dependent attempt frequency is considered the agreement between experimental data and numerical calculations is excellent while, when we assume a constant attempt frequency is also in very good agreement, but under estimate the experimental value. Negative enhancement factor as observed $Al$ solvent, this could promote an enhancement of the solvent diffusion coefficient for less diluted alloys. 

Finally, the $F=1$ and $F\neq 1$ approximations yield practically to the same results for the $L_{ij}$ and $D^{\star}$ in both systems here studied. Differences were observed for the ratios $D^{\star}_B/D^{\star}_A$, $L_{AB}/L_{BB}$, $f_B$ and $b_{A^{\star}}$, evidently not reflected in the self and solute diffusion coefficients, despite the notorious differences observed for the strong/weak attractive interaction between the solute $U/Al$-vacancy diluted in $Al,Ni$ hosts respectively.

The vacancy tracer diffusion coefficient for the $NiAl$ and $AlU$ system were compared with available experimental data obtaining an excellent agreement with the here described theory. Calculations for the diffusion coefficient of the paired specie, shows that a vacancy drag mechanism could occur for $AlU$ when $F\neq 1$, but is unlikely to occur for $NiAl$ in both, $F=1$ and $F\neq 1$. 

This opens the door for future works in the same direction where similar procedure will be used that includes interstitial defects.

\section*{Acknowledgements}
I am grateful to A.M.F. Rivas and Joaqu\'{\i}n Guill\'en, for comments on the manuscript. Also, I am grateful to Mart\'in Urtubey for the Figure \ref{FIG2}. This work was partially financed by CONICET PIP-00965/2010 and the CNEA/CAC - Gerencia Materiales. 

\end{document}